\documentclass[a4paper]{jpconf}
\usepackage{graphicx}
 \usepackage[T1]{fontenc}
 \usepackage{bm}

\newcommand{\be}{\begin{equation}}
\newcommand{\ee}{\end{equation}}
\newcommand{\ba}{\begin{array}}
\newcommand{\ea}{\end{array}}
\newcommand{\bt}{\begin{tabular}}
\newcommand{\et}{\end{tabular}}
\newcommand{\bn}{\begin{eqnarray}}
\newcommand{\en}{\end{eqnarray}}
\newcommand{\bns}{\begin{eqnarray*}}
\newcommand{\ens}{\end{eqnarray*}}

\newcommand{\stext}[1]{\mbox{\rm{\scriptsize{#1}}}}
\renewcommand{\vec}[1]{{\mathbf{#1}}}

\newcommand{\half}{{1\over 2}}

\newcommand{\disregard}[1]{}

\begin{document}
\title{Current Developments in Nuclear Density Functional Methods}

\author{Jacek Dobaczewski}

\address{Institute of Theoretical Physics, University of Warsaw,
             Ho\.za 69, PL-00681 Warsaw, Poland}
\address{Department of Physics, P.O. Box 35 (YFL),
             FI-40014 University of Jyv\"askyl\"a, Finland}

\ead{Jacek.Dobaczewski@fuw.edu.pl}

\begin{abstract}
Density functional theory (DFT) became a universal approach to
compute ground-state and excited configurations of many-electron
systems held together by an external one-body potential in
condensed-matter, atomic, and molecular physics. At present, the DFT
strategy is also intensely studied and applied in the area of nuclear
structure. The nuclear DFT, a natural extension of the
self-consistent mean-field theory, is a tool of choice for
computations of ground-state properties and low-lying excitations of
medium-mass and heavy nuclei. Over the past thirty-odd years, a lot
of experience was accumulated in implementing, adjusting, and using
the density-functional methods in nuclei. This research
direction is still extremely actively pursued. In
particular, current developments concentrate on (i) attempts to
improve the performance and precision delivered by the nuclear
density-functional methods, (ii) derivations of density functionals
from first principles rooted in the low-energy chromodynamics and
effective theories, and (iii) including effects of low-energy
correlations and symmetry restoration. In this study, we present
an overview of recent results and achievements gained in nuclear
density-functional methods.
\end{abstract}

\section{Introduction}

The Density Functional Theory (DFT) was introduced in atomic physics
through the Hohenberg-Kohn \cite{[Hoh64a]} and Kohn-Sham
\cite{[Koh65a]} theorems. Its quantum-mechanical foundation relies on
a simple variational concept that uses observables as variational
parameters. Namely, for any Hamiltonian $\hat{H}$ and observable $\hat{Q}$,
one can formulate the constraint variational problem,
\be\label{eq:01}
\delta \langle \hat{H} - \lambda \hat{Q} \rangle = 0 ,
\ee
whereby the total energy of the system $E\equiv \langle \hat{H}
\rangle$ becomes a function of the observable $Q\equiv \langle
\hat{Q} \rangle$, that is $E = E(Q)$, provided the Lagrange multiplier
$\lambda$ can be eliminated from functions $E(\lambda)$ and
$Q(\lambda)$ that are obtained from the variation in
Eq.~(\ref{eq:01}) performed at fixed $\lambda$.
This can be understood in terms of a two-step variational procedure.
First, Eq.~(\ref{eq:01}) ensures that the total energy is minimized
at fixed $Q$, and second, the minimization of $E(Q)$ in function of
$Q$ gives obviously the exact ground-state energy $E_0=\min_Q E(Q)$
and the exact value $Q_0$ of the observable $Q$ calculated for the
ground-state wave function.

Function $E(Q)$ is thus the simplest model of the density functional.
However, the idea of two-step variational procedure can be applied to
an arbitrary observable or a set of observables, and hence the total
energy can become a function $E(Q_k)$ of several observables $Q_k$,
$\delta \langle \hat{H} - \sum_k\lambda_k \hat{Q}_k \rangle = 0
\Longrightarrow E = E(Q_k)$,
or a functional $E[Q(q)]$ of a continuous set of observables $Q(q)$,
$\delta \langle \hat{H} - \int\rmd q\,\lambda(q) \hat{Q}(q) \rangle = 0
\Longrightarrow E = E[Q(q)]$.
\disregard{
\be\label{eq:02}
\delta \langle \hat{H} - \sum_k\lambda_k \hat{Q}_k \rangle = 0
~~~~\Longrightarrow~~~~ E = E(Q_k) ,
\ee
or a functional $E[Q(q)]$ of a continuous set of observables $Q(q)$,
\be\label{eq:03}
\delta \langle \hat{H} - \int\rmd q\,\lambda(q) \hat{Q}(q) \rangle = 0
~~~~\Longrightarrow~~~~ E = E[Q(q)] .
\ee
}

When these ideas are applied to the observable
$\hat{\rho}(\vec{r}) = \sum_{i=1}^A \delta(\vec{r}-\vec{r}_i)$, which is the local
density of a many-body system at point $\vec{r}$, we obtain the original local DFT,
\be\label{eq:04}
\delta \langle \hat{H} - \int\rmd \vec{r}\,V(\vec{r}) \hat{\rho}(\vec{r}) \rangle = 0
~~~~\Longrightarrow~~~~ E = E[\rho(\vec{r})] ,
\ee
whereby the local external potential $V(\vec{r})$ plays the role of the Lagrange
multiplier that selects a given density profile ${\rho}(\vec{r})$. By the
same token, the nonlocal DFT is obtained by using a nonlocal external
potential $V(\vec{r},\vec{r}\,')$,
\be\label{eq:05}
\delta \langle \hat{H} - \int\rmd \vec{r}\int\rmd \vec{r}\,'\,V(\vec{r},\vec{r}\,') \hat{\rho}(\vec{r},\vec{r}\,') \rangle = 0
~~~~\Longrightarrow~~~~  E = E[\rho(\vec{r},\vec{r}\,')].
\ee
In each of these cases, by minimizing the functionals $E[\rho(\vec{r})]$
or $E[\rho(\vec{r},\vec{r}\,')]$, we obtain the exact ground-state
energy of the many-body system along with its exact local ${\rho}(\vec{r})$
or nonlocal $\rho(\vec{r},\vec{r}\,')$ one-body density.

It is thus obvious that the idea of two-step variational principle,
which is at the heart of DFT, does not give us any hint on which
observable has to be picked as the variational parameter. Moreover,
the exact derivation of the density functional is entirely
impractical, because it involves solving exactly the variational
problem that is equivalent to finding the exact ground state. If we
were capable of doing that, no DFT would have been further required.
Nevertheless, the exact arguments presented above can serve us as a
justification of {\it modelling} the ground-state properties of
many-body systems by DFT, which, however, must be rather guided by
physical intuition, general theoretical arguments, experiment, and
exact calculations for simple systems.

\section{Fundamentals}

\begin{figure}[h]
\begin{center}\vspace*{-4mm}
\begin{minipage}[b]{0.42\textwidth}
\begin{center}
\includegraphics[width=\textwidth]{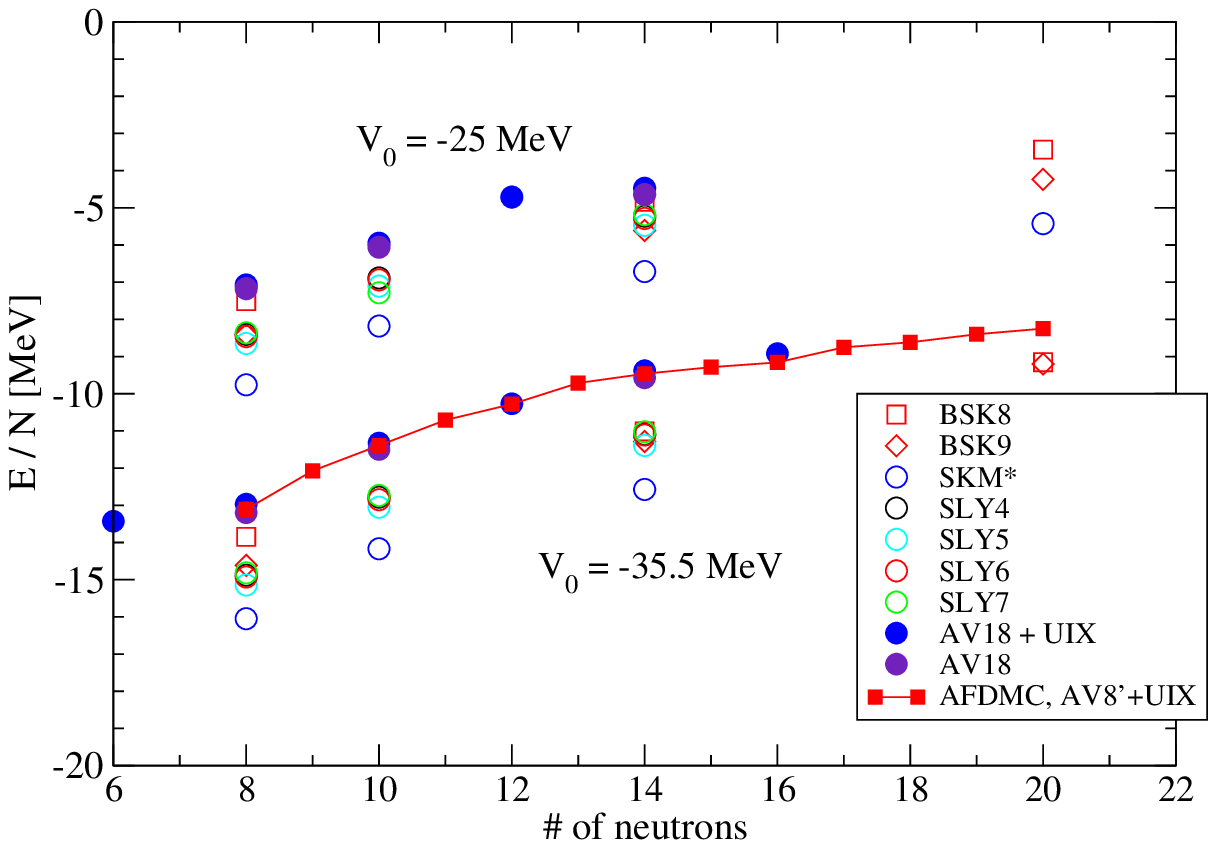}
\caption{\label{INPC2010.100708-16-05}Energies of neutron drops
confined by the Woods-Saxon potential, see the text. From Ref.~\cite{[Wir10]};
picture courtesy of R.B.\ Wiringa.}
\end{center}
\end{minipage}\hspace{0.01\textwidth}%
\begin{minipage}[b]{0.56\textwidth}
In practice, the exact density functionals (\protect\ref{eq:04}) or (\protect\ref{eq:05})
are modelled as integrals of energy densities ${\cal H}$, and thus they are
called energy-density functionals (EDFs). They can be local functions
of local densities (\protect\ref{eq:06}), quasilocal functions of
local higher-order densities (\protect\ref{eq:07}), nonlocal functions of local densities
(\protect\ref{eq:08}), or nonlocal functions of nonlocal densities (\protect\ref{eq:09}).
\bn\label{eq:06}
E[\rho(\vec{r})] &\!\!\!\!=\!\!\!\!& \!\!\int\!\!\rmd \vec{r}\,{\cal H}(\rho(\vec{r})),
\\
\label{eq:07}
E[\rho(\vec{r})] &\!\!\!\!=\!\!\!\!&
\!\!\int\!\!\rmd \vec{r}\,{\cal H}(\rho(\vec{r}),\tau(\vec{r}),
\Delta\rho(\vec{r}),\ldots),
\\
\label{eq:08}
E[\rho(\vec{r})] &\!\!\!\!=\!\!\!\!& \!\!\int\!\!\rmd \vec{r}\!\!\int\!\!\rmd \vec{r}\,'\,
{\cal H}(\rho(\vec{r}),\rho(\vec{r}\,')),
\\
\label{eq:09}
 E[\rho(\vec{r},\vec{r}\,')] &\!\!\!\!=\!\!\!\!& \!\!\int\!\!\rmd
\vec{r}\!\!\int\!\!\rmd \vec{r}\,'\,{\cal H}(\rho(\vec{r},\vec{r}\,')).
\en
\end{minipage}
\disregard{
\begin{minipage}[b]{0.48\textwidth}
\be\label{eq:06}
E[\rho(\vec{r})] = \int\rmd \vec{r}\,{\cal H}(\rho(\vec{r}))
\ee
\be\label{eq:08}
E[\rho(\vec{r})] =
\int\rmd \vec{r}\,{\cal H}(\rho(\vec{r}),\tau(\vec{r}),
\Delta\rho(\vec{r}),\ldots)
\ee
\be\label{eq:09}
E[\rho(\vec{r})] = \int\rmd \vec{r}\int\rmd \vec{r}\,'\,
{\cal H}(\rho(\vec{r}),\rho(\vec{r}\,'))
\ee
\be\label{eq:10}
 E[\rho(\vec{r},\vec{r}\,')] = \int\rmd \vec{r}\int\rmd \vec{r}\,'\,{\cal H}(\rho(\vec{r},\vec{r}\,'))
\ee
\end{minipage}
\begin{minipage}[b]{0.48\textwidth}
\begin{center}
\includegraphics[width=\textwidth]{INPC2010.100708-16-10.eps}
\caption{\label{INPC2010.100708-16-10}Figure caption. From Ref.~\cite{[Geb10]};
picture courtesy of B.\ Gebremariam.}
\end{center}
\end{minipage}
}
\end{center}
\end{figure}

An example that highlights connection with {\it ab-initio} theory is shown
in Fig.~\ref{INPC2010.100708-16-05}, Ref.~\cite{[Wir10]}, where
binding energies calculated exactly for neutron drops confined within
the Woods-Saxon potential of two different depths, $V_0=-25$ and
$-35.5$\,MeV, are compared with those corresponding to several Skyrme
functionals \cite{[Ben03]}. Calculations of this type, performed at
different depths, surface thicknesses, and deformations of the
confining potential may allow for a better determination of
Skyrme-functional parameters.

Two major classes of approach that are currently used and
developed in nuclear structure physics are based on relativistic and
nonrelativistic EDFs \cite{[Rin96],[Ben03],[Lal04a]}.
The nonrelativistic EDFs are most often built as:
\be\label{eq:10}
{\cal H}(\rho(\vec{r},\vec{r}\,')) =
\half\sum_{vt} \left(
\hat{V}^{\stext{dir}}_{vt}(\vec{r},\vec{r}\,')
\left[\left[\rho^t_v(\vec{r})\rho^t_v(\vec{r}\,')\right]^0\right]_0
-\hat{V}^{\stext{exc}}_{vt}(\vec{r},\vec{r}\,')
\left[\left[\rho^t_v(\vec{r},\vec{r}\,')\rho^t_v(\vec{r}\,',\vec{r})\right]^0\right]_0\right) ,
\ee
where
$\hat{V}^{\stext{dir}}_{vt}$ and $\hat{V}^{\stext{exc}}_{vt}$ denote
the EDF-generating (pseudo)potentials in the direct and exchange
channels, respectively, and we sum over the spin-rank $v=0$ and 1
(scalar and vector) and isospin-rank $t=0$ and 1 (isoscalar and
isovector) spherical-tensor densities \cite{[Car08e],[Car10]}
coupled to the total isoscalar (superscript 0) and scalar (subscript
0) term. For example, finite-range momentum-independent central potentials
generate the Gogny \cite{[Dec80]} or M3Y \cite{[Nak03]} nonlocal
functionals (\ref{eq:10}) and zero-range momentum-dependent
pseudopotentials generate the Skyrme \cite{[Ben03]} or BCP
\cite{[Bal08]} quasilocal functionals (\ref{eq:08}).

Expression (\ref{eq:10}) derives from the Hartree-Fock formula for
the average energy of a Slater determinant. However, the
EDF-generating pseudopotentials should not be confused with the
nucleon-nucleon (NN) bare or effective interaction or Brueckner
G-matrix. Indeed, their characteristic features are different -- they
neither are meant to describe the NN scattering properties, as the
bare NN force is, nor are meant to be used in a restricted phase
space, as the effective interaction is, nor depend on energy, as the
G-matrix does. Moreover, to ensure correct saturation properties, the
EDF-generating pseudopotentials must themselves depend on the
density. But most importantly, the generated EDFs are modelled so as
to describe the exact binding energies and not those in the
Hartree-Fock approximation, which otherwise would have required
adding higher-order corrections based on the many-body perturbation
theory.

\begin{figure}[h]
\begin{center}
\includegraphics[width=0.59\textwidth]{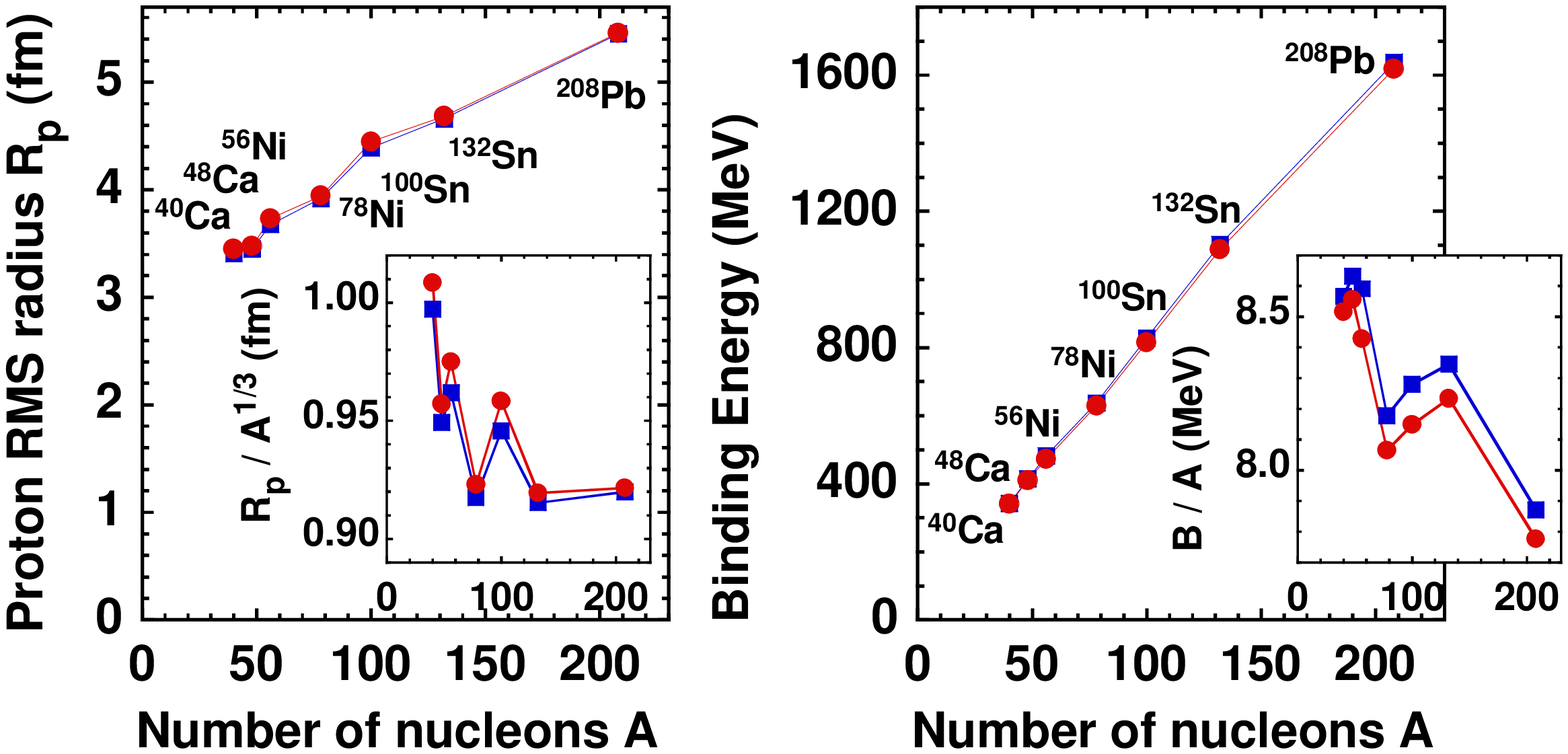}\hspace{0.03\textwidth}%
\begin{minipage}[b]{0.37\textwidth}\caption{\label{INPC2010.100708-16-09}
Proton RMS radii (left)
and binding energies (right) of doubly magic nuclei, determined by using
the Gogny D1S EDF \cite{[Del10]} (squares) and second-order
Skyrme-like EDF \cite{[Dob10]} derived therefrom (circles).
The insets show results in expanded scales and scaled by particle numbers $A$.
}
\end{minipage}
\end{center}
\end{figure}

\begin{figure}
\begin{center}
\includegraphics[width=0.38\textwidth]{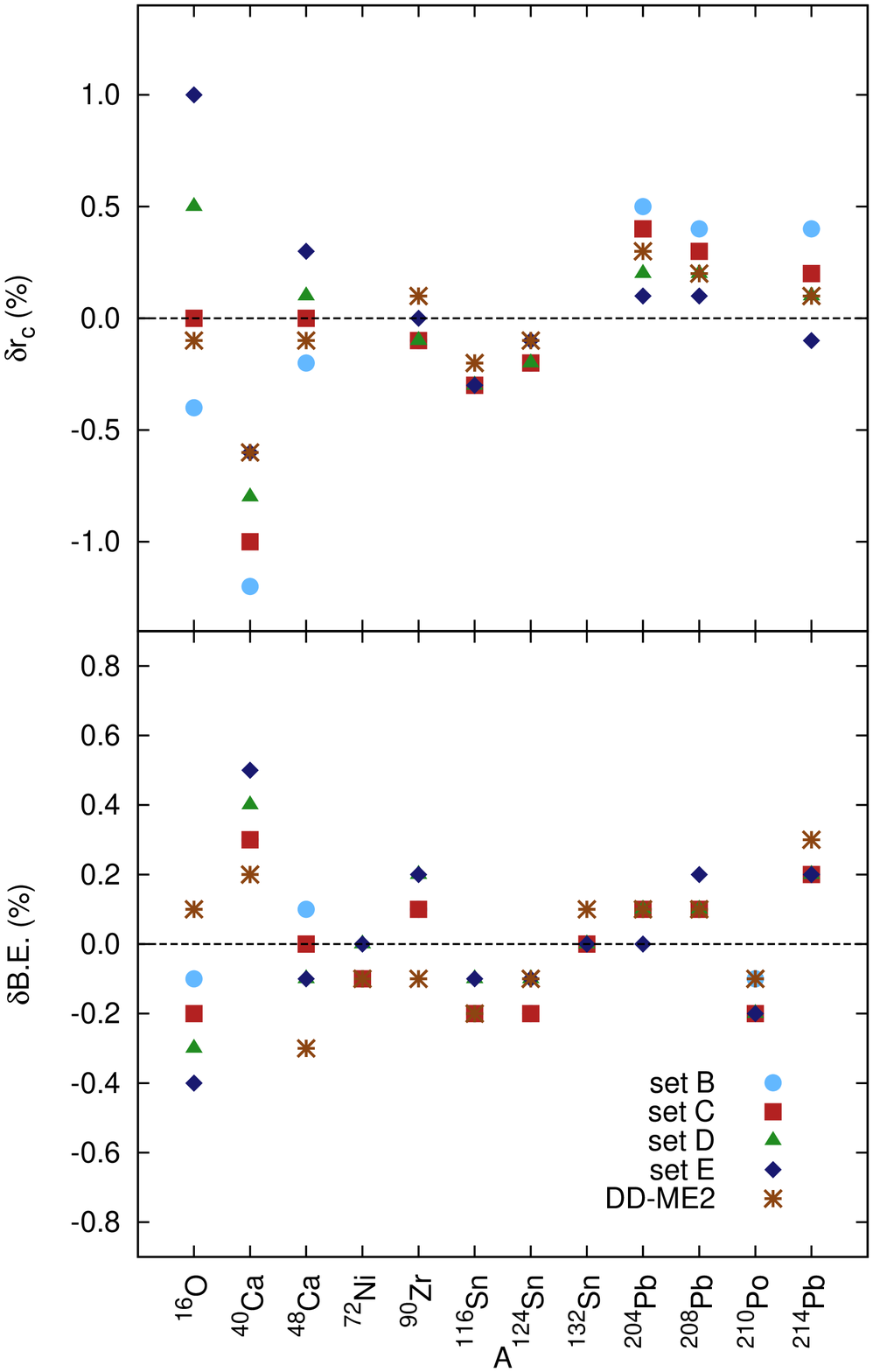}\hspace{0.03\textwidth}%
\begin{minipage}[b]{0.58\textwidth}\caption{\label{Niksic}
The relative deviations (in percentages) between
the experimental and theoretical charge radii (upper panel) and
binding energies (lower panel) of 12 spherical nuclei, calculated with
the meson-exchange interaction DD-ME2 and the four point-coupling
parameter sets. In all cases, the relative deviations are below 1\%.
Starting from the meson-exchange density-dependent interaction
DD-ME2, the equivalent point-coupling
parametrization for the effective Lagrangian was derived by incorporating the
density dependence of the parameters.
The parameters of the point-coupling model were adjusted to reproduce
the nuclear matter equation of state obtained with the DD-ME2
interaction and surface thickness and surface energy of semi-infinite
nuclear matter.
Reprinted figure with permission from Ref.~\cite{[Nik08]}.
Copyright 2008 by the American Physical Society.}
\end{minipage}
\end{center}
\end{figure}

Only very recently, it has been demonstrated \cite{[Dob10],[Car10]}
that the nuclear nonlocal EDFs, based on sufficiently short-range
EDF-generating pseudopotentials, are equivalent to quasilocal EDFs.
In Fig.~\ref{INPC2010.100708-16-09} are compared the proton RMS radii
and binding energies of doubly magic nuclei, determined by using the
Gogny D1S EDF \cite{[Del10]} and second-order Skyrme-like EDF S1Sb
\cite{[Dob10]} derived therefrom by using the Negele-Vautherin (NV)
density-matrix expansion (DME) \cite{[Neg72]}. One can see that
already at second order, the DME gives excellent precision of the order
of 1\%.
In Ref.~\cite{[Nik08]}, similar conclusions were also reached
when comparing the nonlocal and quasilocal relativistic EDFs, see Fig.~\ref{Niksic}.

\begin{figure}
\begin{center}
\begin{minipage}[t]{0.47\textwidth}
\begin{center}
\includegraphics[width=0.7\textwidth]{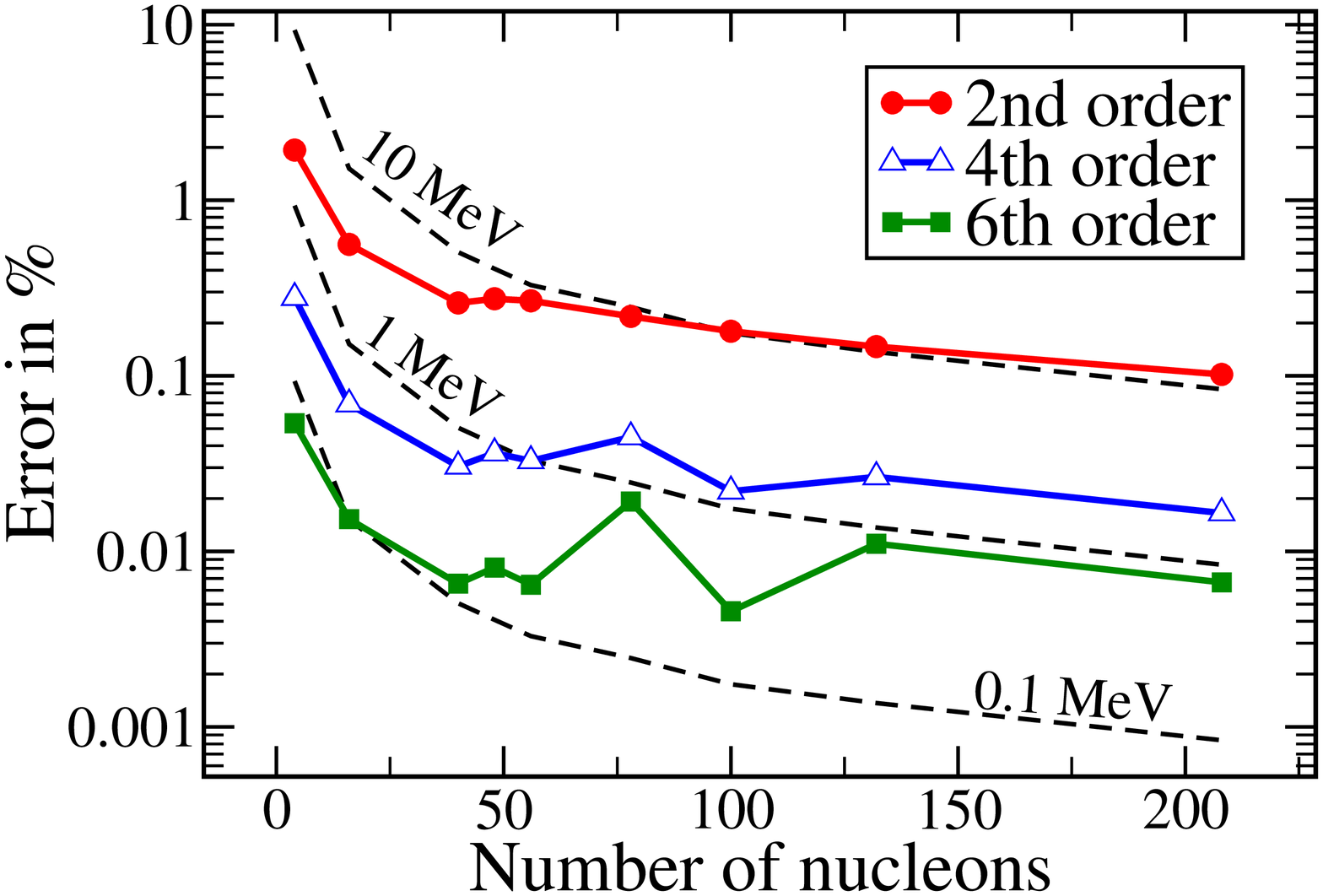}
\caption{\label{DME1}
Precision of the Taylor DME of
the direct interaction energies calculated for the nonlocal Gogny D1S EDF
\cite{[Ber91b]}. The nine nuclei used for the test are
$^4$He, $^{16}$O, $^{40,48}$Ca, $^{56,78}$Ni, $^{100,132}$Sn, and
$^{208}$Pb.
From Ref.~\cite{[Car10]}.}
\end{center}
\end{minipage}\hspace{0.03\textwidth}%
\begin{minipage}[t]{0.49\textwidth}
\begin{center}
\includegraphics[width=0.7\textwidth]{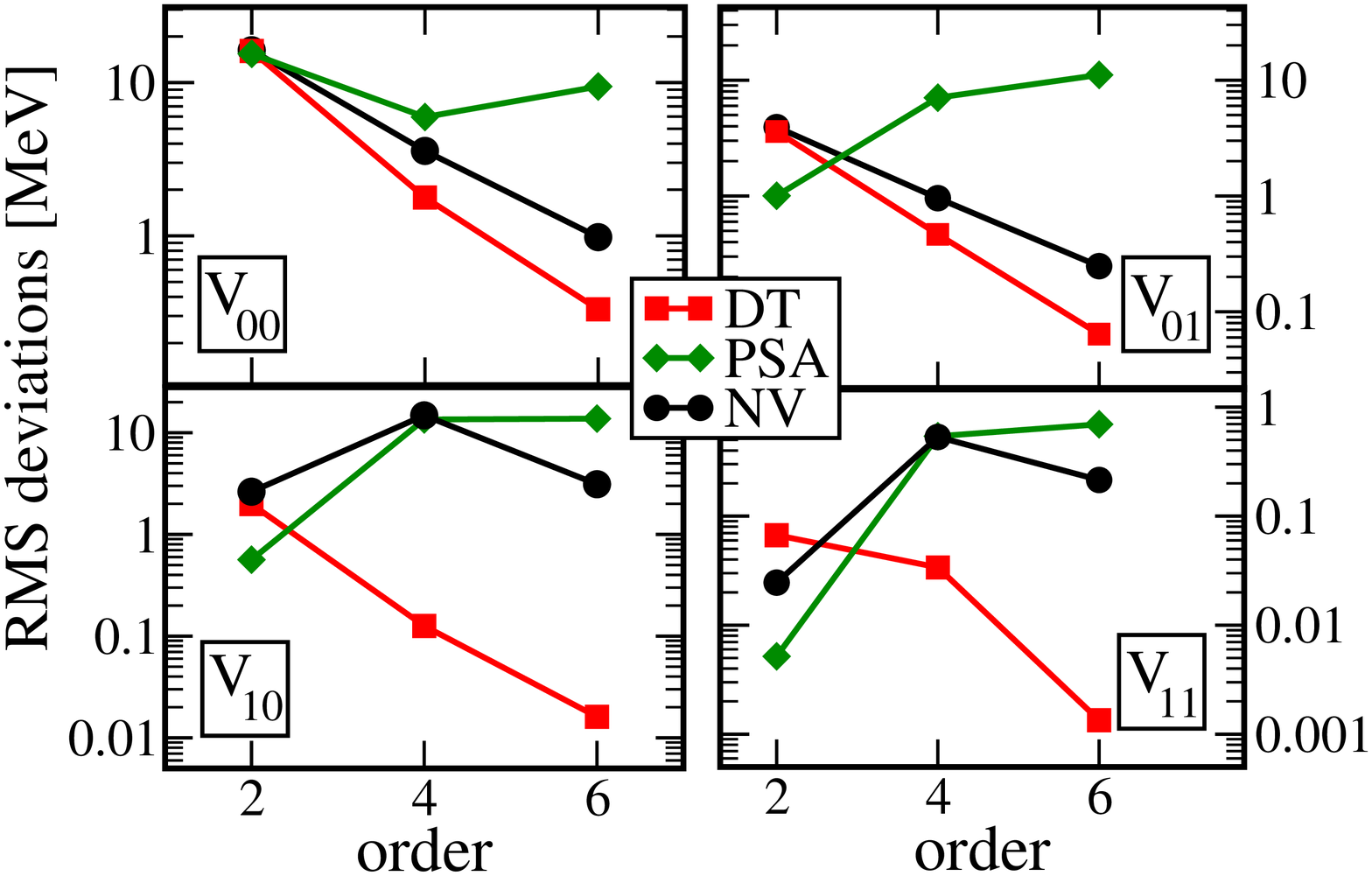}
\caption{\label{DME2}
The RMS deviations between the exact and approximate exchange
energies calculated for the nine
nuclei listed in the caption of Fig.~\protect\ref{DME1}, see the text.
From Ref.~\cite{[Car10]}.}
\end{center}
\end{minipage}
\end{center}
\end{figure}

Figs.~\ref{DME1} and \ref{DME2} show convergence
of the direct and exchange interaction energies, respectively, when
the Taylor and damped Taylor (DT) DMEs are performed up to sixth
order \cite{[Car10]}. The four panels of Fig.~\ref{DME2} show results
obtained in the four spin-isospin channels labeled by $V_{vt}$.
Results of the DT DME \cite{[Car10]} are compared with those
corresponding to the NV \protect\cite{[Neg72]} and PSA
\protect\cite{[Geb10]} expansions. It is extremely gratifying to see
that in each higher order the precision increases by a large factor,
which is characteristic to a rapid power-law convergence.
The success and convergence of the DME expansions relies on the fact
that the finite-range nuclear effective interactions are very
short-range as compared to the spatial variations of nuclear
densities. The quasilocal (gradient) expansion in nuclei works!

\section{Applications}

\begin{figure}
\begin{center}
\begin{minipage}[b]{0.48\textwidth}
\begin{center}
\includegraphics[height=0.6\textwidth]{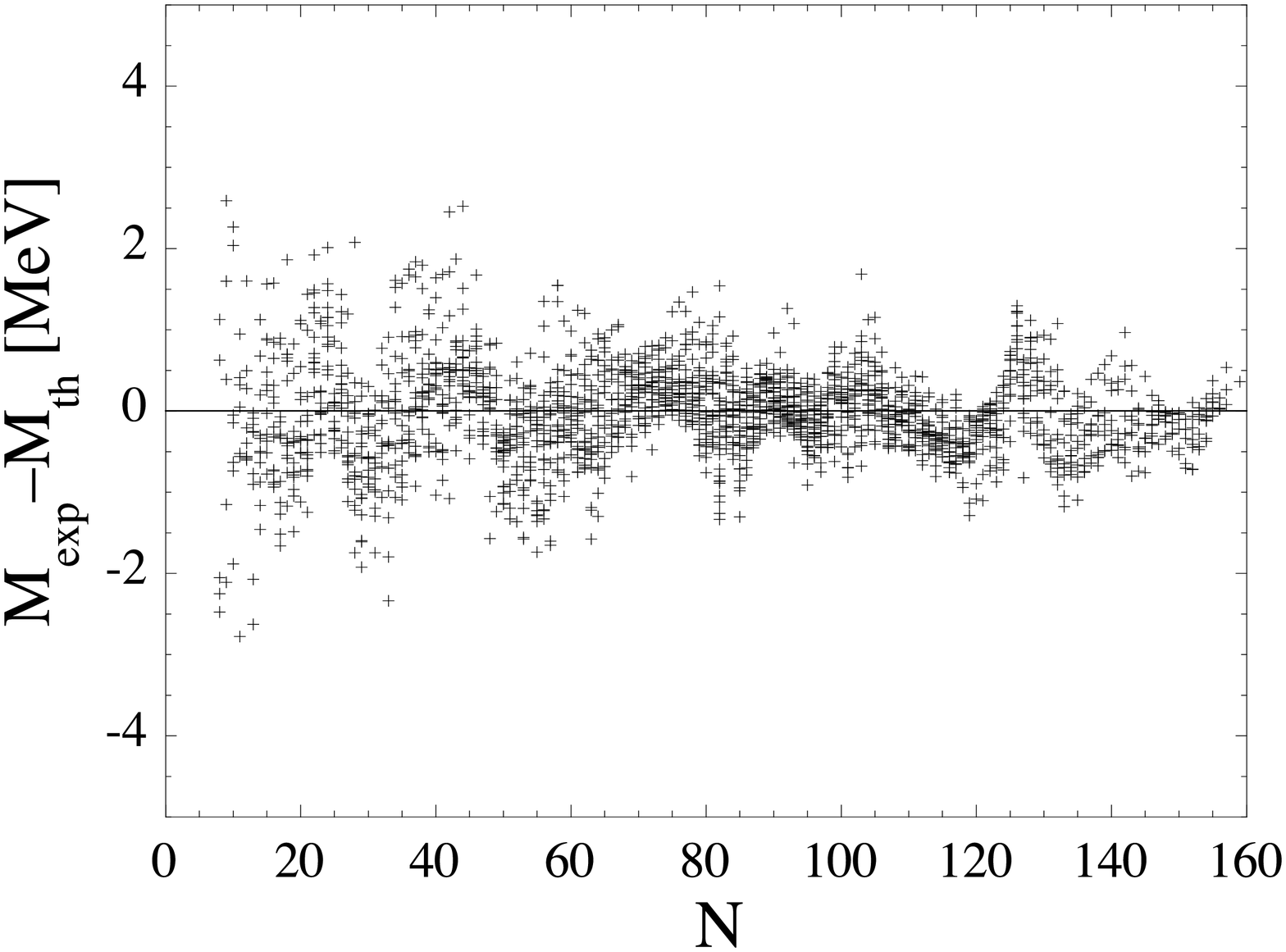}
\caption{\label{Goriely2}
Differences between measured \cite{[Aud03]} and HFB-17 \cite{[Gor09]} masses,
as a function of the neutron number $N$.
Reprinted figure with permission from Ref.~\cite{[Gor09]}.
Copyright 2009 by the American Physical Society.}
\end{center}
\end{minipage}\hspace{0.03\textwidth}%
\begin{minipage}[b]{0.48\textwidth}
\begin{center}
\includegraphics[height=0.6\textwidth]{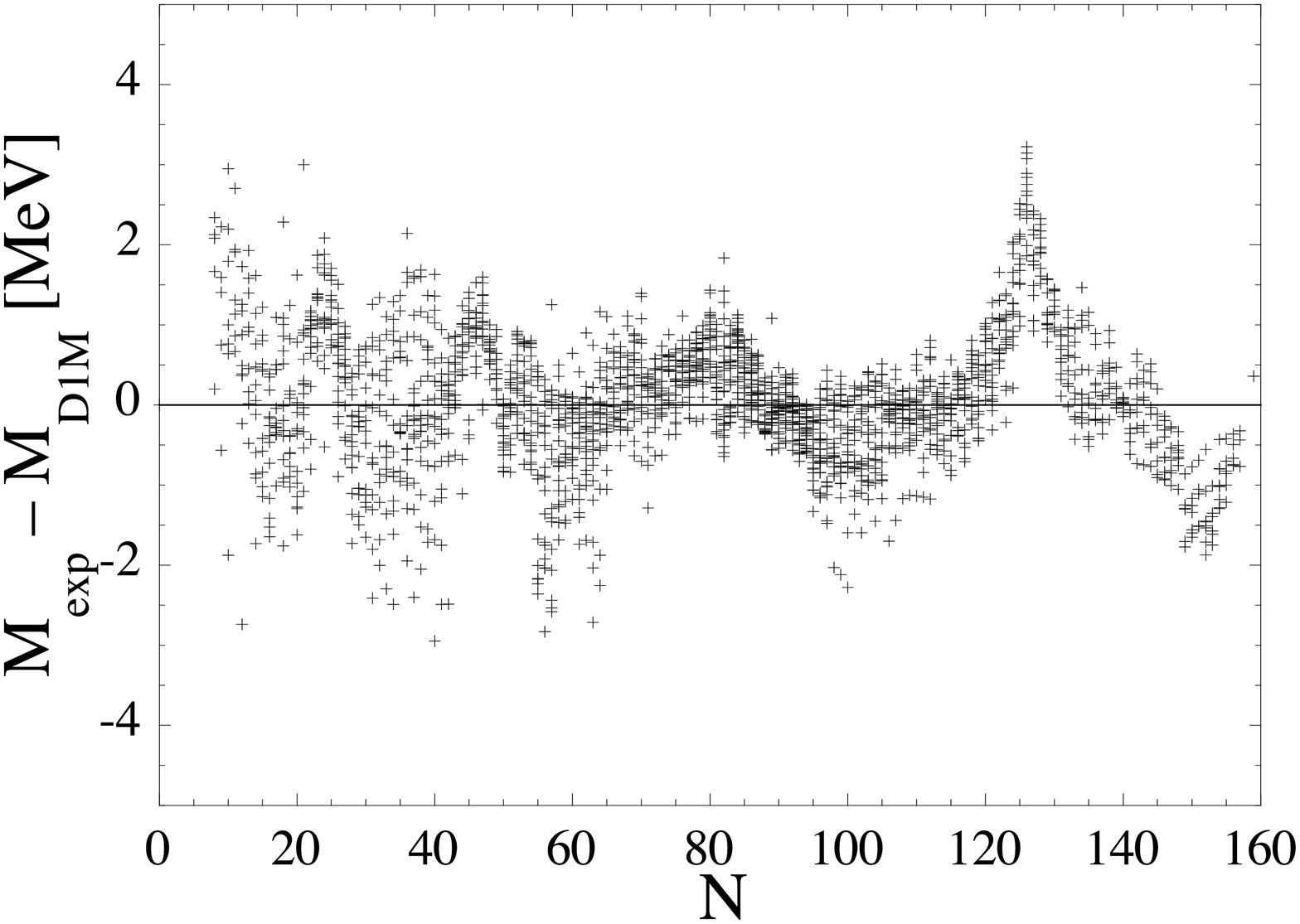}
\caption{\label{Goriely1}
Differences between measured \cite{[Aud03]} and D1M \cite{[Gor09a]} masses,
as a function of the neutron number $N$.
Reprinted figure with permission from Ref.~\cite{[Gor09a]}.
Copyright 2009 by the American Physical Society.}
\end{center}
\end{minipage}
\end{center}
\end{figure}

When compared to the experimental binding energies, the quasilocal
Skyrme functional HFB-17 \cite{[Gor09]} (Fig.~\ref{Goriely2}) gives
results, which have the quality very similar to those given by the
nonlocal Gogny functional D1M \cite{[Gor09a]} (Fig.~\ref{Goriely1}).
In both cases, the functionals were augmented by terms responsible
for the pairing correlations and all parameters were adjusted
specifically to binding energies. Moreover, in both cases, by using
either the 5D collective Hamiltonian approach or configuration
mixing, theoretical binding energies were corrected for collective
quadrupole correlations. The results are truly impressive, with the
RMS deviations calculated for 2149 masses being as small as 798 and
581\,keV, respectively.

The problem of treating collective correlations and excitations
within the DFT or EDF approaches is one of the most important issues
currently studied in applications to nuclear systems. The question of
whether one can describe these effects by using the functional only
is not yet resolved. In practice, relatively simple functionals that
are currently in use, require adding low-energy correlation effects
explicitly. This can be done by reverting from the description
in terms of one-body densities back to the wave functions of mean-field
states. For example, for quadrupole correlations this amounts to using
the following configuration-mixing states,
\be\label{eq:11}
|\Psi_{NZ,JM}\rangle = \int \rmd\beta\rmd\gamma \sum_K f_K(\beta,\gamma)
\hat{P}_N \hat{P}_Z \hat{P}_{JMK} |\Psi(\beta,\gamma) \rangle ,
\ee
where $\hat{P}_N$, $\hat{P}_Z$, and $\hat{P}_{JMK}$ are projection
operators on good neutron number $N$, proton number $Z$, and angular
momentum $J$ with laboratory and intrinsic projections $M$ and $K$.
The intrinsic mean-field wave functions $|\Psi(\beta,\gamma) \rangle$
correspond to one-body densities constrained to quadrupole
deformations $\beta$ and $\gamma$ \cite{[RS80]}.

To determine variationally
the mixing amplitudes $f_K(\beta,\gamma)$, one has to generalize the energy
densities, such as those shown in Eqs.~(\ref{eq:06})--(\ref{eq:09}),
to transition energy densities that
enable us to compute Hamiltonian kernels.
For mean-field
states, this can be rigorously done by using the Wick theorem, whereby the
average energy $\langle\Psi|\hat{H}|\Psi\rangle$ generalizes to
matrix element $\langle\Psi_1|\hat{H}|\Psi_2\rangle$ as \cite{[RS80]}:
\bn\label{eq:12}
 \langle\Psi|\hat{H}|\Psi\rangle &\simeq&
\int\rmd \vec{r}\int\rmd \vec{r}\,'\,{\cal H}(\rho(\vec{r},\vec{r}\,'))
 ~~~~\mbox{for}~~\rho(\vec{r},\vec{r}\,')
={\displaystyle\frac{\langle\Psi|a^+(\vec{r}\,')a(\vec{r}\,')|\Psi\rangle}
      {\langle\Psi                             |\Psi\rangle}} ,
\\  \label{eq:13}
   \langle\Psi_1|\hat{H}|\Psi_2\rangle &\simeq&
\int\rmd \vec{r}\int\rmd \vec{r}\,'\,{\cal H}(\rho_{12}(\vec{r},\vec{r}\,'))
 ~~~~\mbox{for}~~\rho_{12}(\vec{r},\vec{r}\,')
={\displaystyle\frac{\langle\Psi_1|a^+(\vec{r}\,')a(\vec{r}\,')|\Psi_2\rangle}
      {\langle\Psi_1                             |\Psi_2\rangle}} .
\en
Although for densities of {\em correlated} states, which are employed
in DFT or EDF methods, this prescription cannot be properly
justified, it has been successfully used in many practical
applications. However, even this simple prescription creates problems
\cite{[Dob07d]}, which may require implementing more complicated schemes
\cite{[Dug09]}.

\begin{figure}
\begin{center}
\begin{minipage}[b]{0.38\textwidth}
\begin{center}
\includegraphics[width=0.8\textwidth]{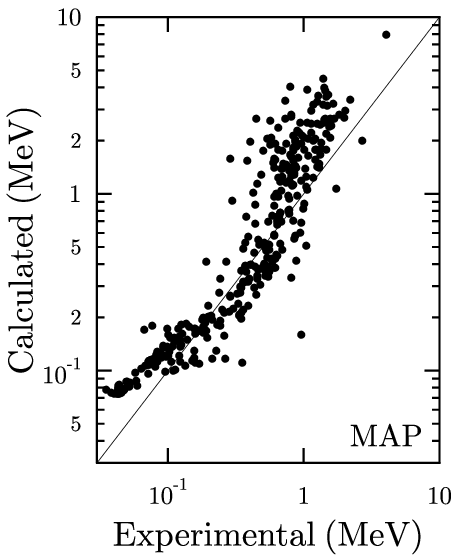}
\caption{\label{sabbey}
Scatter plot comparing the theoretical and experimental
$2^+_1$ excitation energies of the 359 nuclei included in the survey
of Ref.~\cite{[Sab07]};
picture courtesy of M.\ Bender.}
\end{center}
\end{minipage}\hspace{0.03\textwidth}%
\begin{minipage}[b]{0.58\textwidth}
\begin{center}
\includegraphics[width=0.8\textwidth]{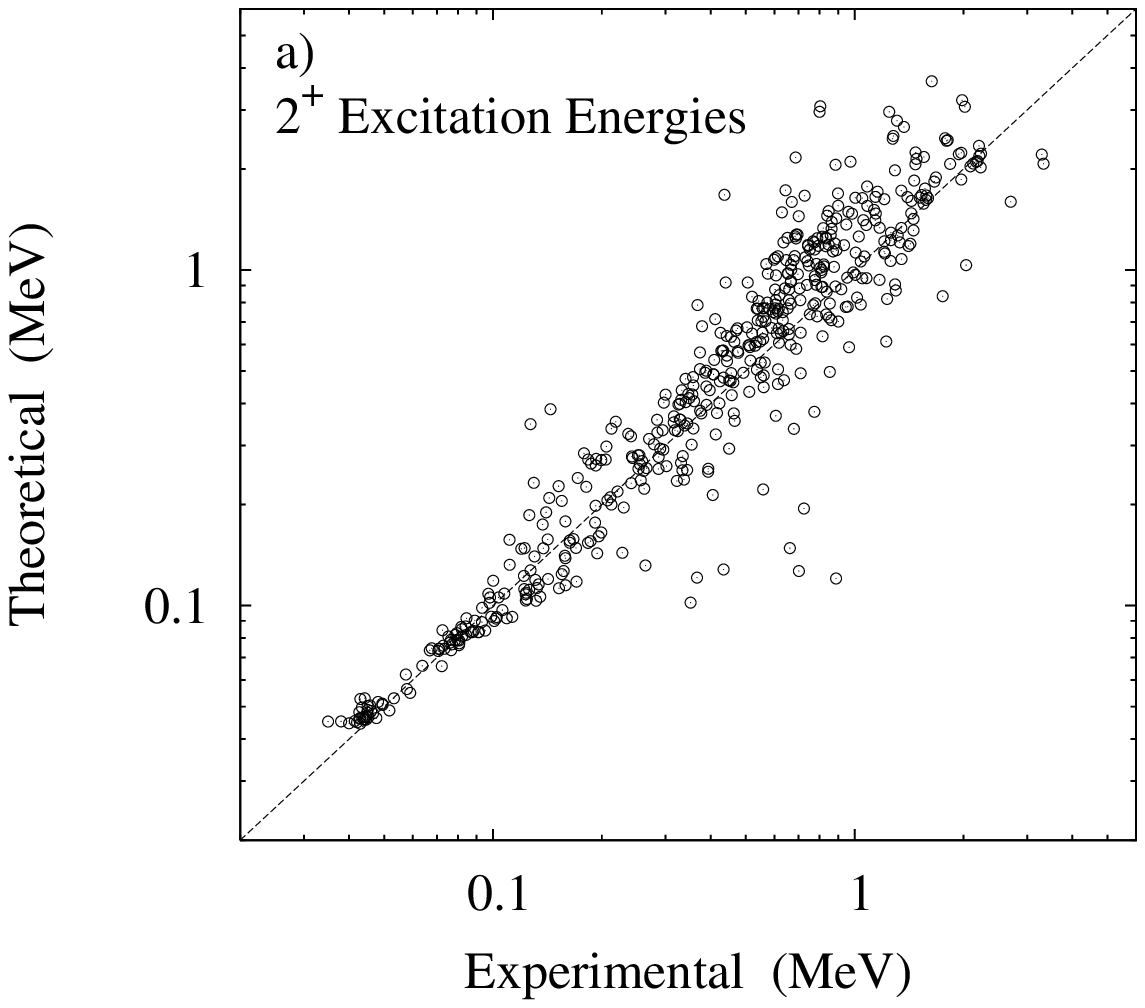}
\caption{\label{e21}
Theoretical $2^+_1$ excitation energies of 537 even-even nuclei
as a function of their experimental values \cite{[Ram01]}.
Reprinted figure with permission from Ref.~\cite{[Del10]}.
Copyright 2010 by the American Physical Society.}
\end{center}
\end{minipage}
\end{center}
\end{figure}

\begin{figure}
\begin{center}
\begin{minipage}[b]{0.54\textwidth}
\begin{center}
\includegraphics[width=0.45\textwidth]{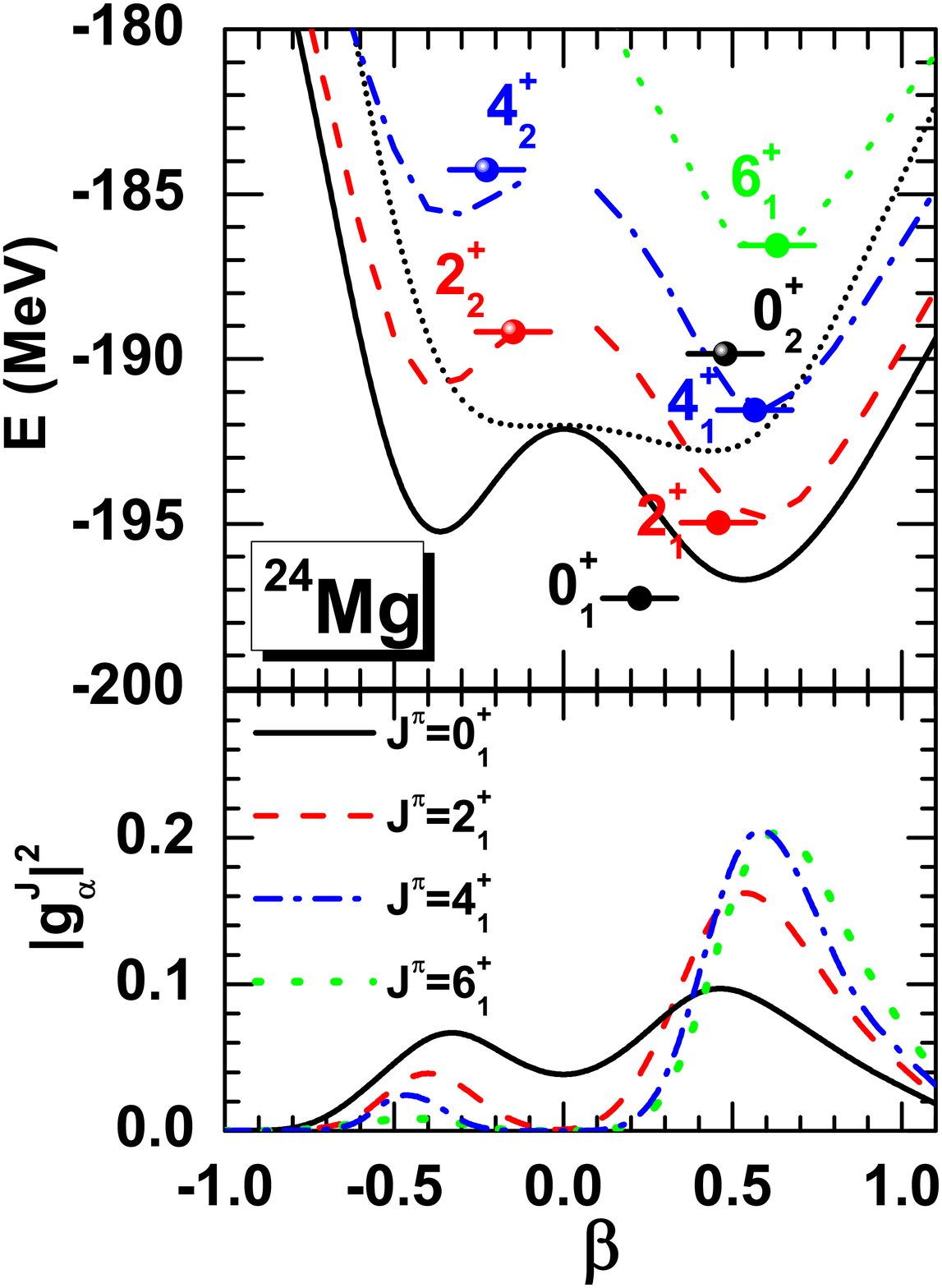}
\caption{\label{Yao}
Energies and the average
axial deformations for the two lowest quadrupole states with angular
momenta 0$^+$, 2$^+$, 4$^+$, and 6$^+$ in $^{24}$Mg, together with the mean-field
(dotted) and the corresponding angular-momentum-projected energy
curves and squares of collective wave functions.
Reprinted figure with permission from Ref.~\cite{[Yao10]}.
Copyright 2010 by the American Physical Society.}
\end{center}
\end{minipage}\hspace{0.03\textwidth}%
\begin{minipage}[b]{0.42\textwidth}
\begin{center}
\includegraphics[width=\textwidth]{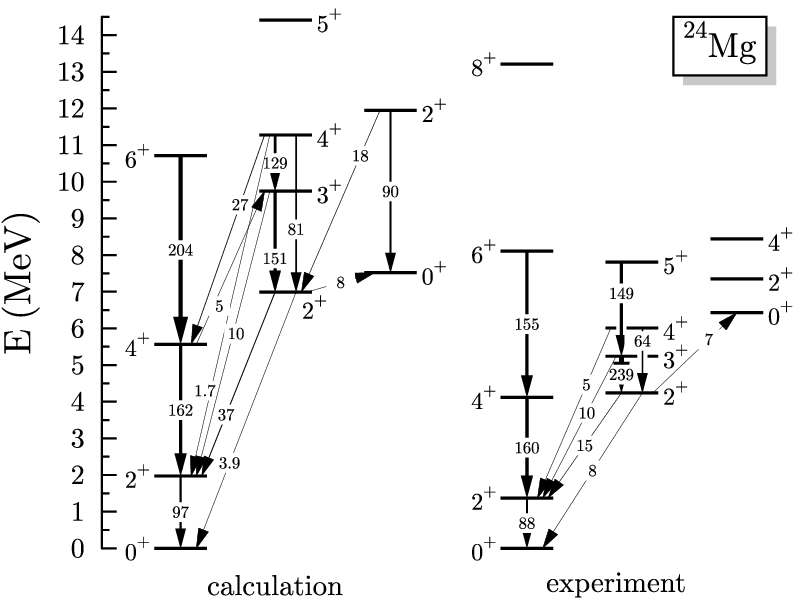}
\caption{\label{Bender}
Excitation spectra and B(E2) values in e$^2$\,fm$^4$ compared to the available
experimental data in $^{24}$Mg. The spectrum is subdivided into
a ground-state band, a $\gamma$ band, and additional low-lying states that
do not necessarily form a band.
From Ref.~\cite{[Ben08a]};
picture courtesy of M.\ Bender.}
\end{center}
\end{minipage}
\end{center}
\end{figure}

An example of the results is shown in Fig.~\ref{sabbey}, where
calculated $2^+_1$ excitation energies \cite{[Sab07]} are compared
with experimental data. One obtains fairly good description for
nuclei across the nuclear chart. Calculations slightly overestimate
the data, which is most probably related to the fact that in this
study the nonrotating mean-field states were used, see
Ref.~\cite{[Zdu08]} and references cited therein. As shown in
Fig.~\ref{e21}, this deficiency disappears when the moments of
inertia of the 5D collective Hamiltonian are determined by using
infinitesimal rotational frequencies \cite{[Del10]}. At present,
calculations using in light nuclei the triaxial projected states of
Eq.~(\ref{eq:11}) are becoming possible for the relativistic
(Fig.~\ref{Yao}) and quasilocal (Fig.~\ref{Bender}) functionals.

\begin{figure}
\begin{center}
\begin{minipage}[b]{0.52\textwidth}
\begin{center}
\includegraphics[width=0.9\textwidth]{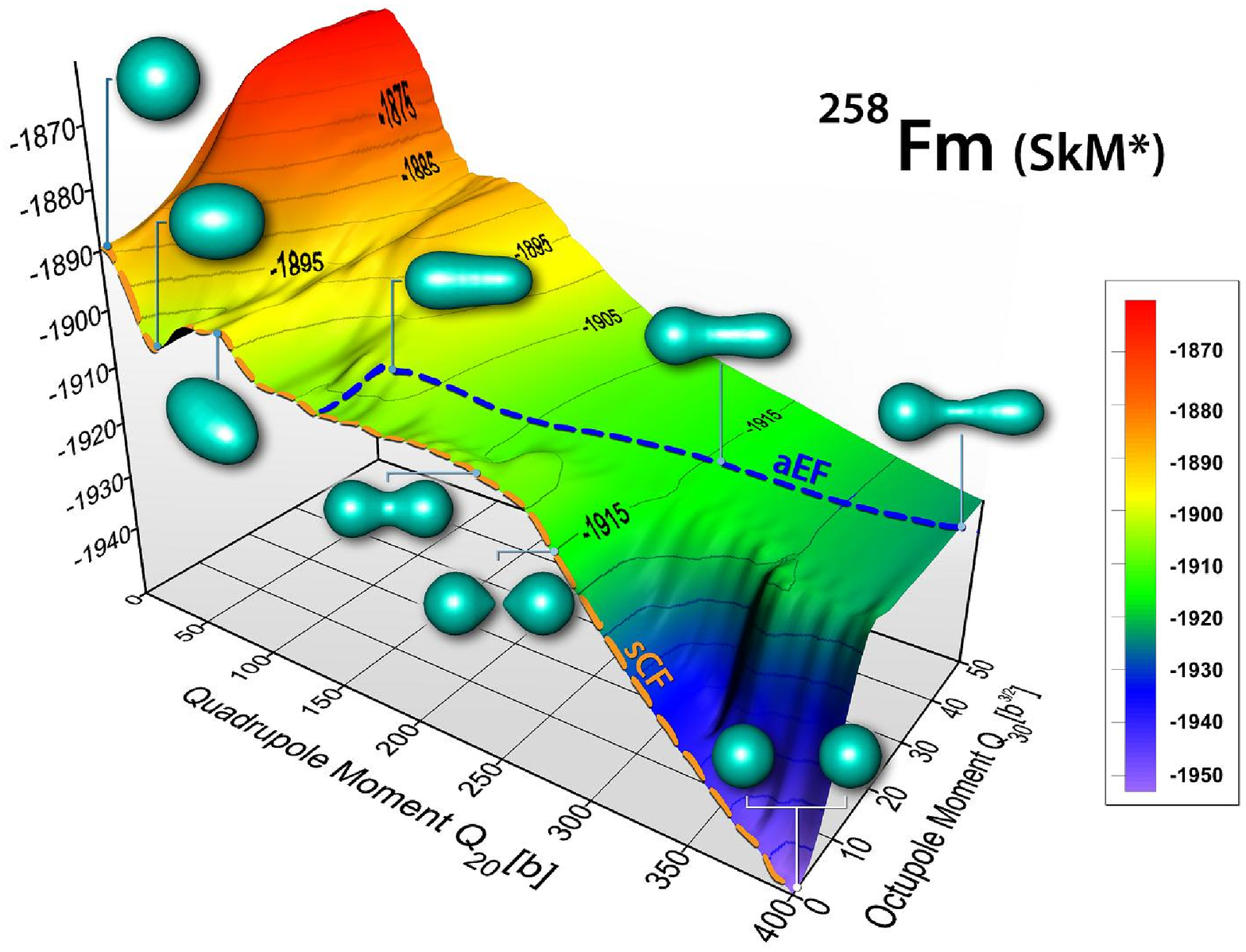}
\caption{\label{Fm258}
The energy surface of $^{258}$Fm calculated for the
quasilocal EDF as a function of two collective variables: the total
quadrupole moment $Q_{20}$ representing the elongation of nuclear shape,
and the total octupole moment $Q_{30}$ representing the left-right shape
asymmetry.
Indicated are the two static fission valleys: asymmetric path aEF
and symmetric-compact path sCF.
From Ref.~\cite{[Ber07a]};
picture courtesy of A.\ Staszczak.}
\end{center}
\end{minipage}\hspace{0.03\textwidth}%
\begin{minipage}[b]{0.44\textwidth}
\begin{center}
\includegraphics[width=0.7\textwidth]{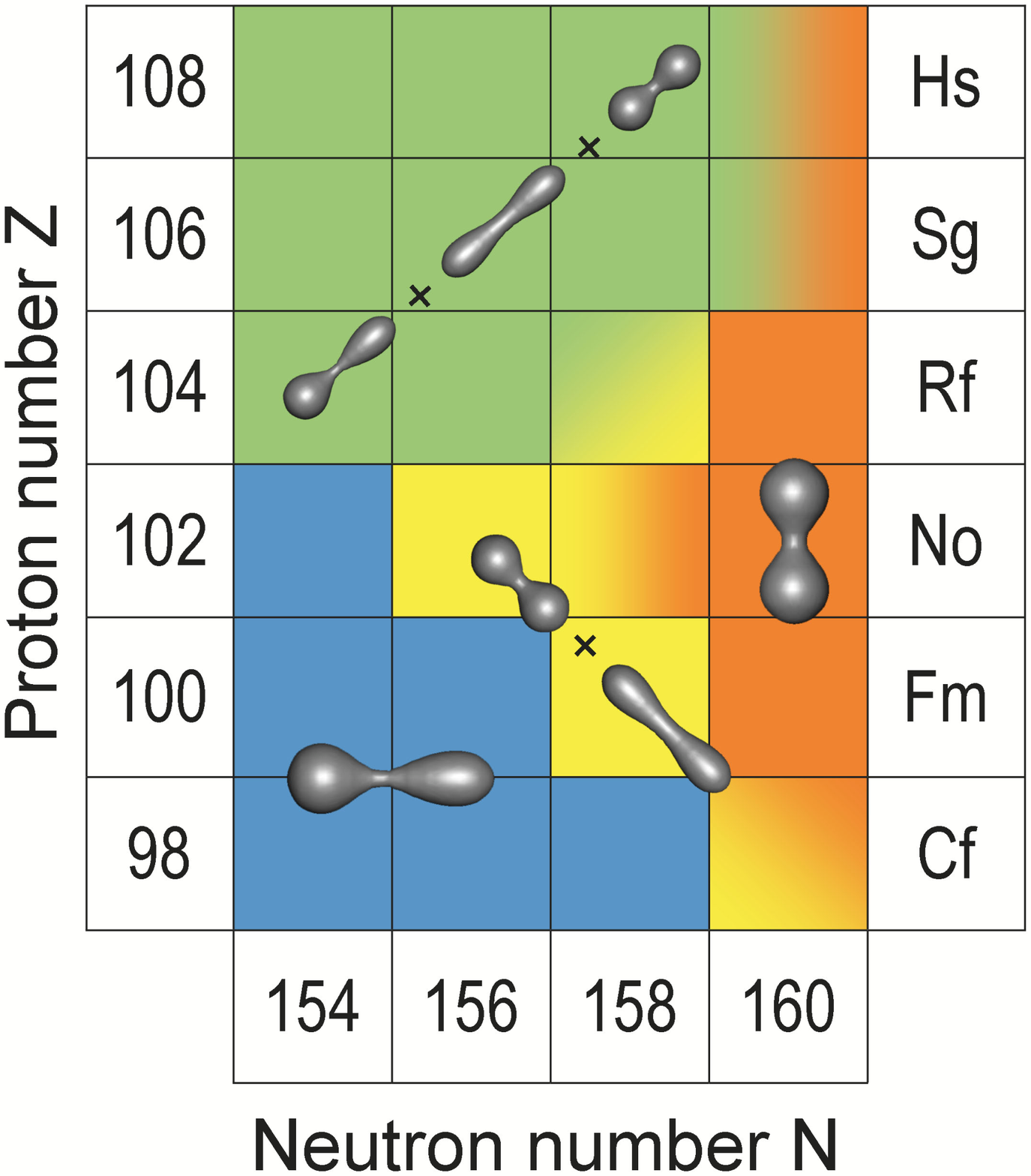}
\caption{\label{bimodal}
Summary of fission pathway results
obtained in Ref.~\cite{[Sta09]}. Nuclei around $^{252}$Cf are predicted to
fission along the asymmetric path and those around $^{262}$No along
the symmetric path. These two regions are separated by
the bimodal symmetric fission around $^{258}$Fm. In a
number of the Rf, Sg, and Hs nuclei, all three fission modes are likely
(trimodal fission). From Ref.~\cite{[Sta09]}}
\end{center}
\end{minipage}
\end{center}
\end{figure}

Another fascinating collective phenomenon that can presently be
described for the non-local \cite{[Gou05],[You09]} and quasilocal
functionals \cite{[Sta09]} is the fission of very heavy nuclei. In
Fig.~\ref{Fm258}, an example of fission-path calculations performed
in $^{258}$Fm is shown in function of the elongation and
shape-asymmetry parameters. One obtains correct description
of the region of nuclei where the phenomenon of bimodal fission occurs
and predicts regions of the trimodal fission, see Fig.~\ref{bimodal}.

\begin{figure}
\begin{center}
\begin{minipage}[b]{0.39\textwidth}
\begin{center}
\includegraphics[width=\textwidth]{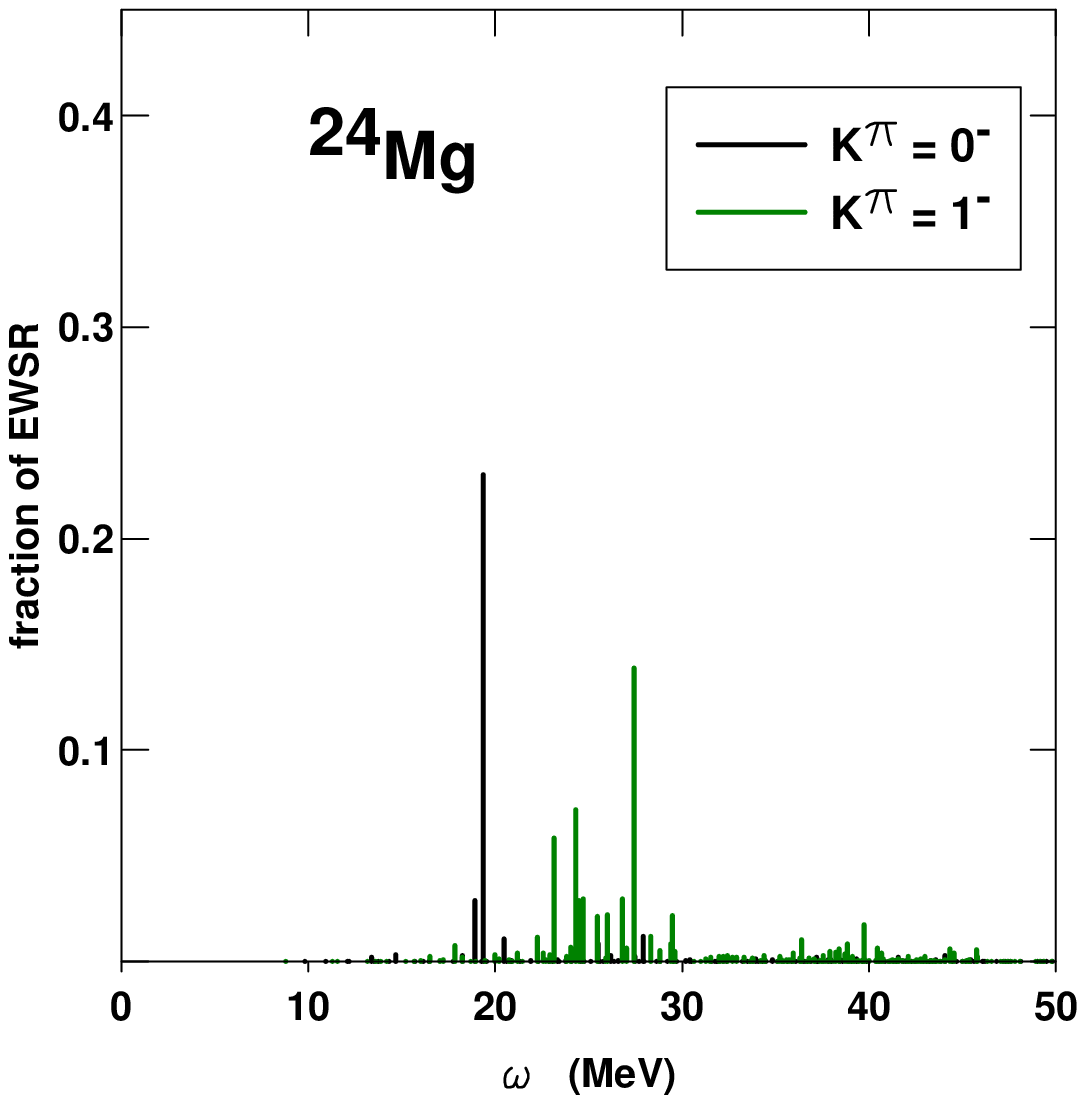}
\caption{\label{Peru}
Dipole excitations in $^{24}$Mg determined for the
nonlocal Gogny D1S functional.
From Ref.~\cite{[Per08a]};
picture courtesy of S.\ P\'eru.}
\end{center}
\end{minipage}\hspace{0.03\textwidth}%
\begin{minipage}[b]{0.57\textwidth}
\begin{center}
\includegraphics[width=\textwidth]{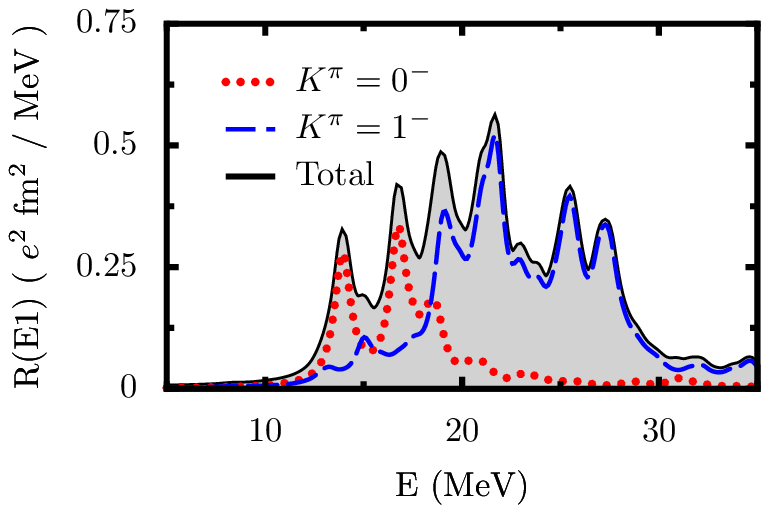}
\caption{\label{Arteaga}
Dipole excitations in $^{20}$Ne determined for the
relativistic NL3 functional.
Reprinted figure with permission from Ref.~\cite{[Art08]}.
Copyright 2008 by the American Physical Society.}
\end{center}
\end{minipage}
\end{center}
\end{figure}

\begin{figure}
\begin{center}
\begin{minipage}[b]{0.55\textwidth}
\begin{center}
\includegraphics[width=0.7\textwidth]{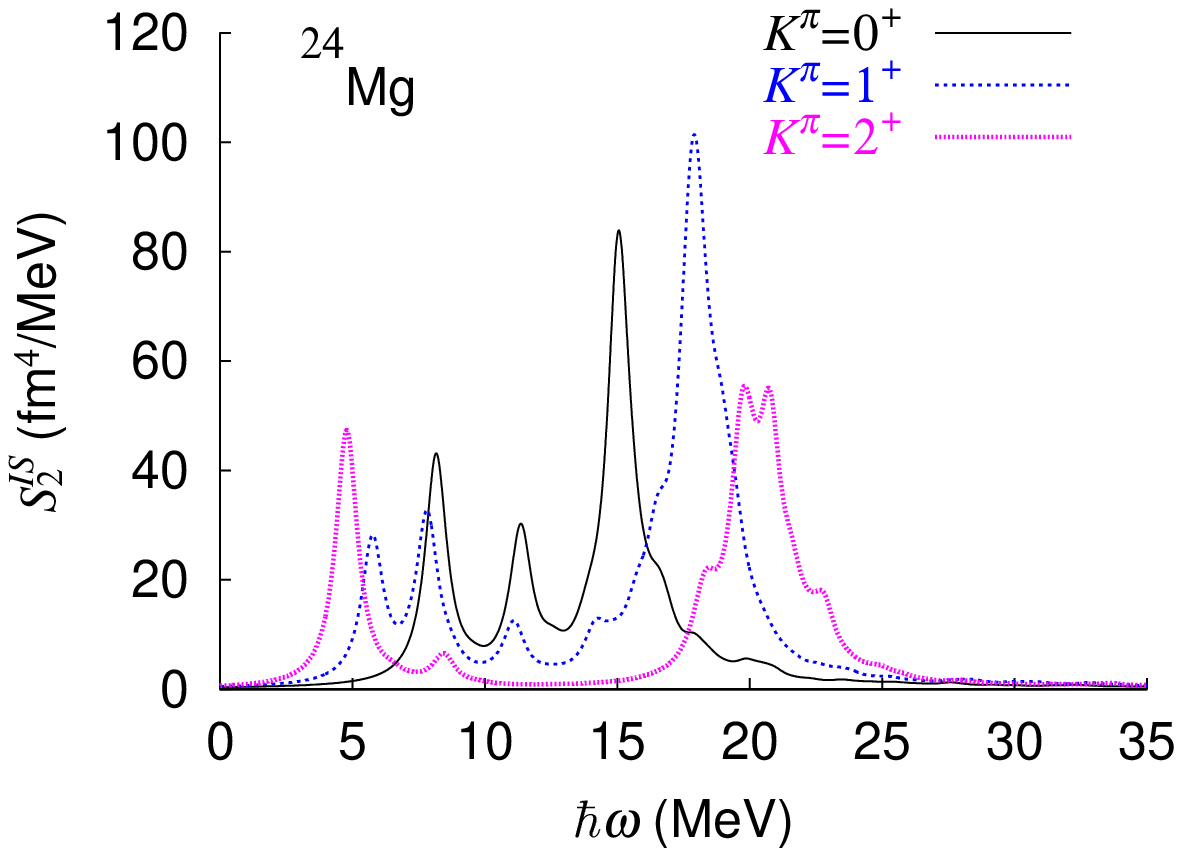}
\caption{\label{Yoshida}
Quadrupole excitations in $^{24}$Mg determined for the
quasilocal Skyrme SkM* functional.
Reprinted figure with permission from Ref.~\cite{[Yos08]}.
Copyright 2008 by the American Physical Society.}
\end{center}
\end{minipage}\hspace{0.03\textwidth}%
\begin{minipage}[b]{0.41\textwidth}
\begin{center}
\includegraphics[width=\textwidth]{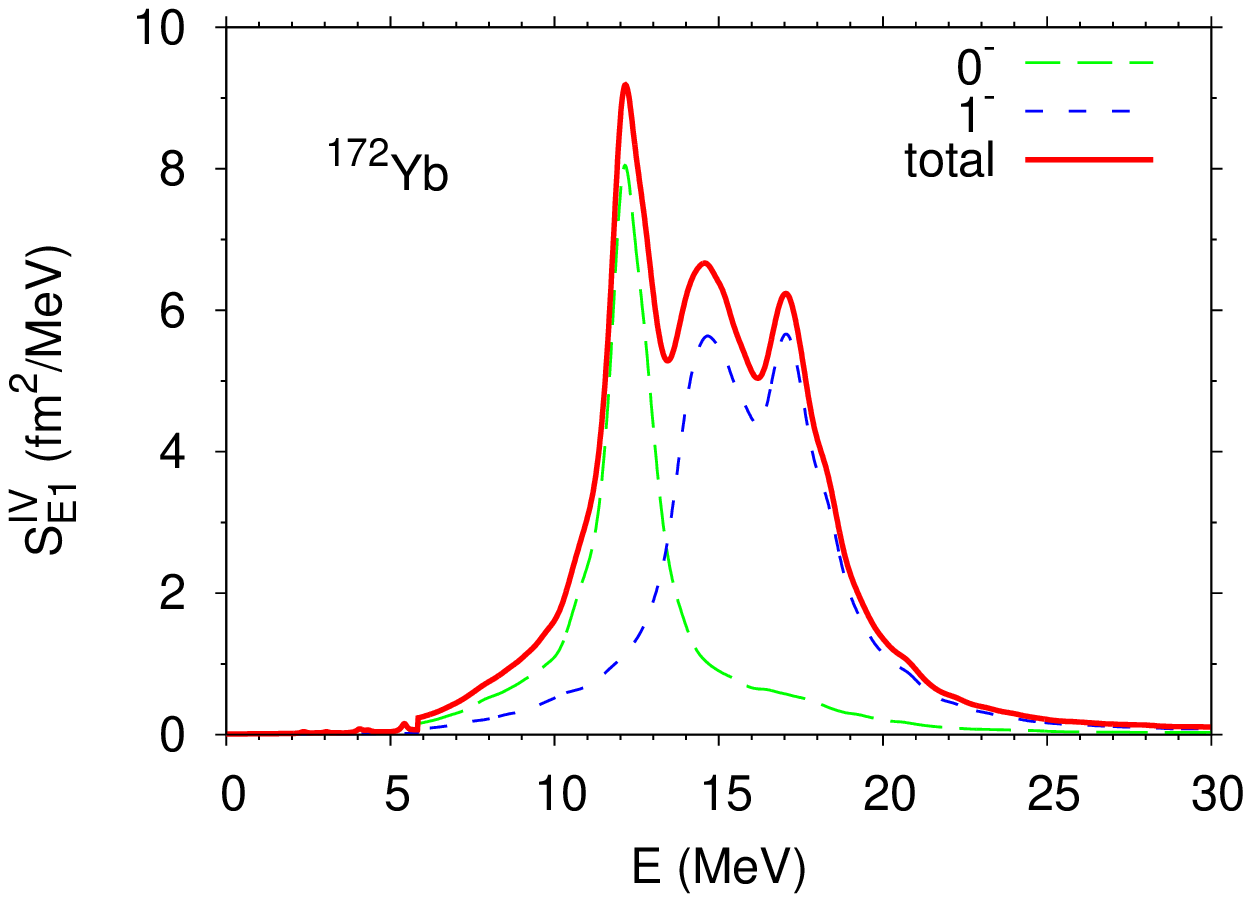}
\caption{\label{Terasaki}
Dipole excitations in $^{172}$Yb determined for the
quasilocal Skyrme SkM* functional.
From Ref.~\cite{[Ter10]};
picture courtesy of J.\ Terasaki.}
\end{center}
\end{minipage}
\end{center}
\end{figure}

In recent years, significant progress was achieved in determining
the multipole giant resonances in deformed nuclei by using the RPA and
QRPA methods. In light nuclei, the multipole modes can be determined
for the nonlocal, relativistic, and quasilocal functionals, see
Figs.~\ref{Peru}, \ref{Arteaga}, and \ref{Yoshida}, respectively. In
heavy nuclei, such calculations are very difficult, because the
number of two-quasiparticle configurations that must be taken into
account grows very fast with the size of the single-particle phase
space. Nevertheless, the first calculation of this kind has already
been reported for $^{172}$Yb, see Fig.~\ref{Terasaki}. The future
developments here will certainly rely on the newly developed
iterative methods of solving the RPA and QRPA equations
\cite{[Nak07a],[Ina09],[Toi10]}.

\begin{figure}
\begin{center}
\begin{minipage}[b]{0.53\textwidth}
\begin{center}
\includegraphics[height=0.23\textwidth]{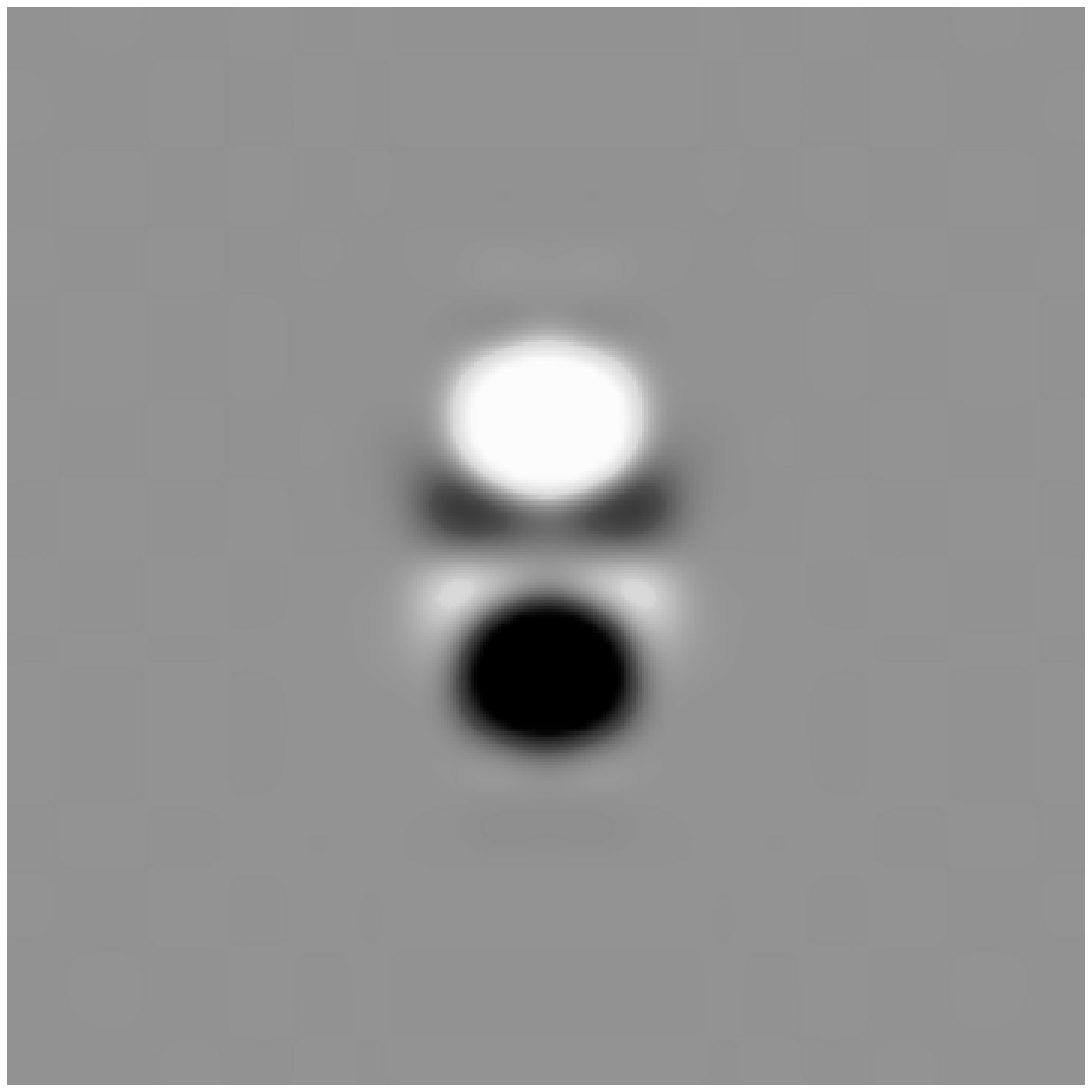}
\includegraphics[height=0.23\textwidth]{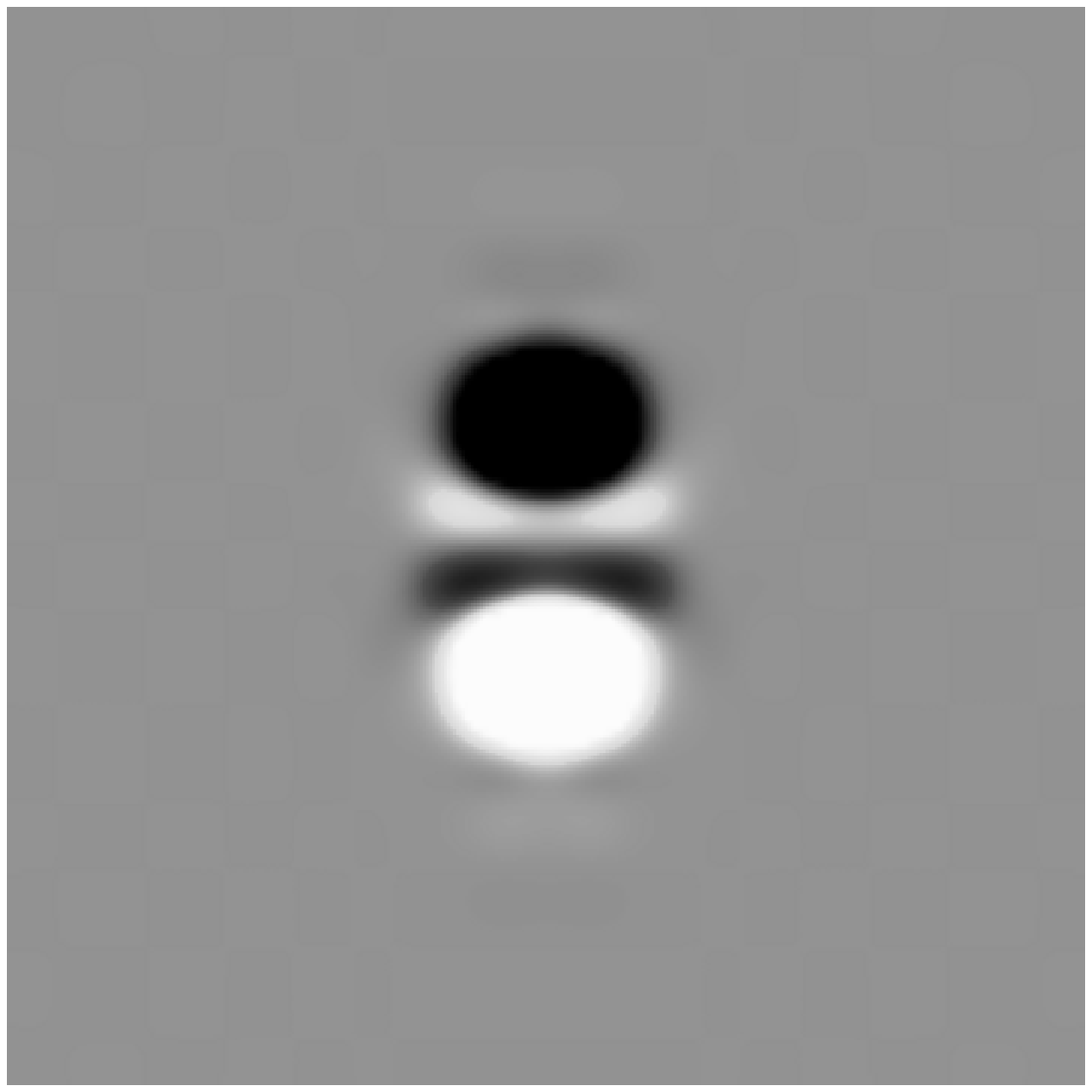}
\includegraphics[height=0.23\textwidth]{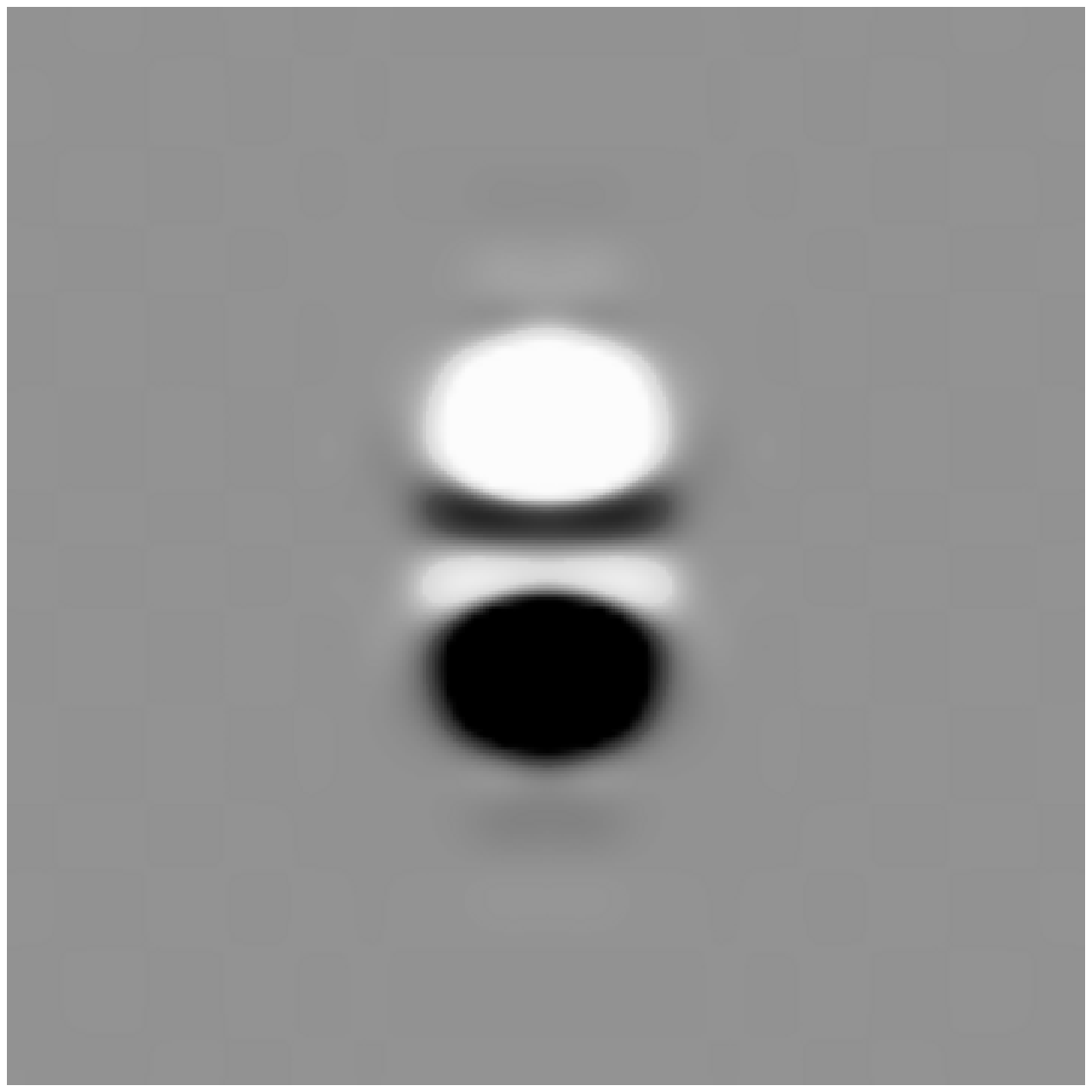}
\includegraphics[height=0.23\textwidth]{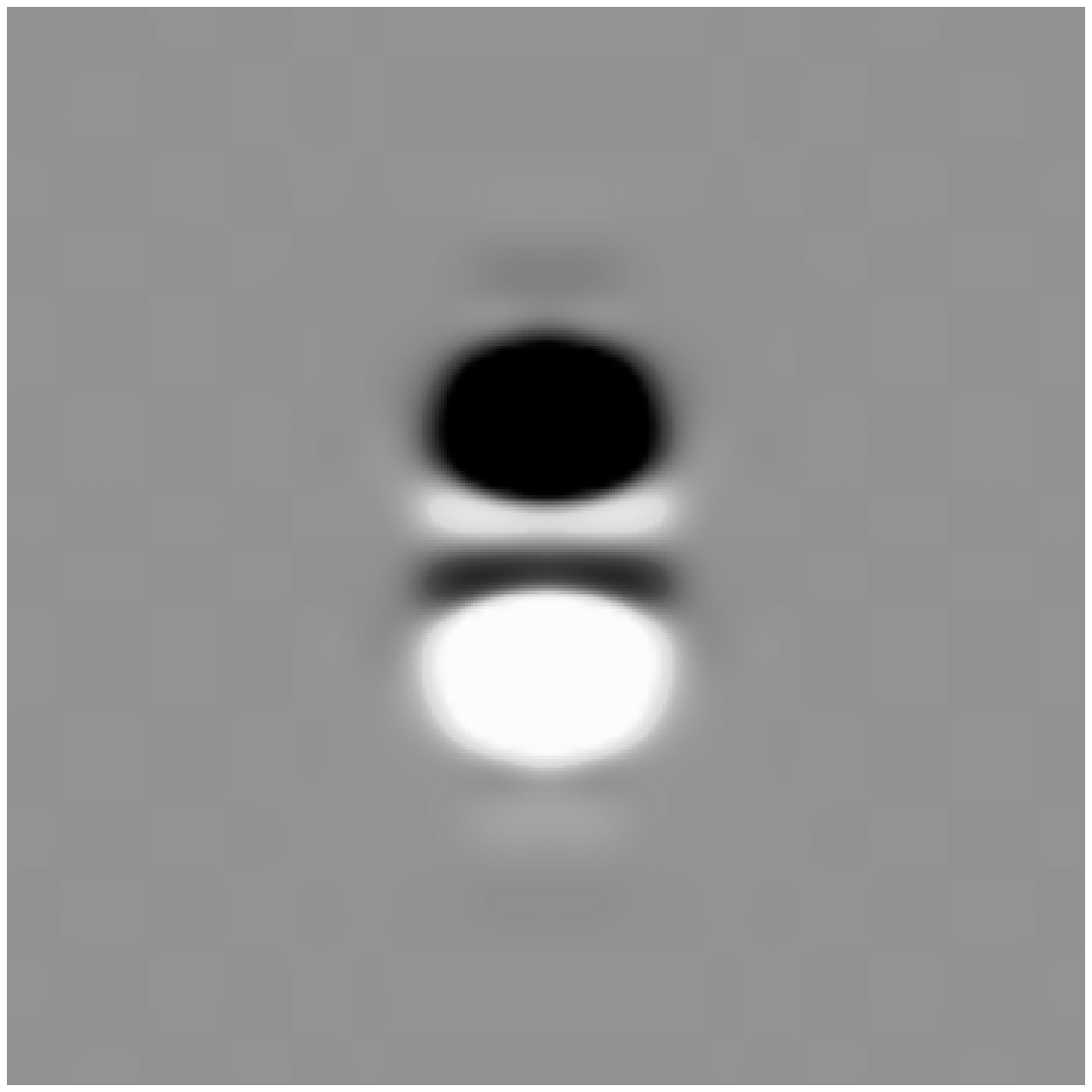}

\vspace*{3mm}
\includegraphics[height=0.23\textwidth]{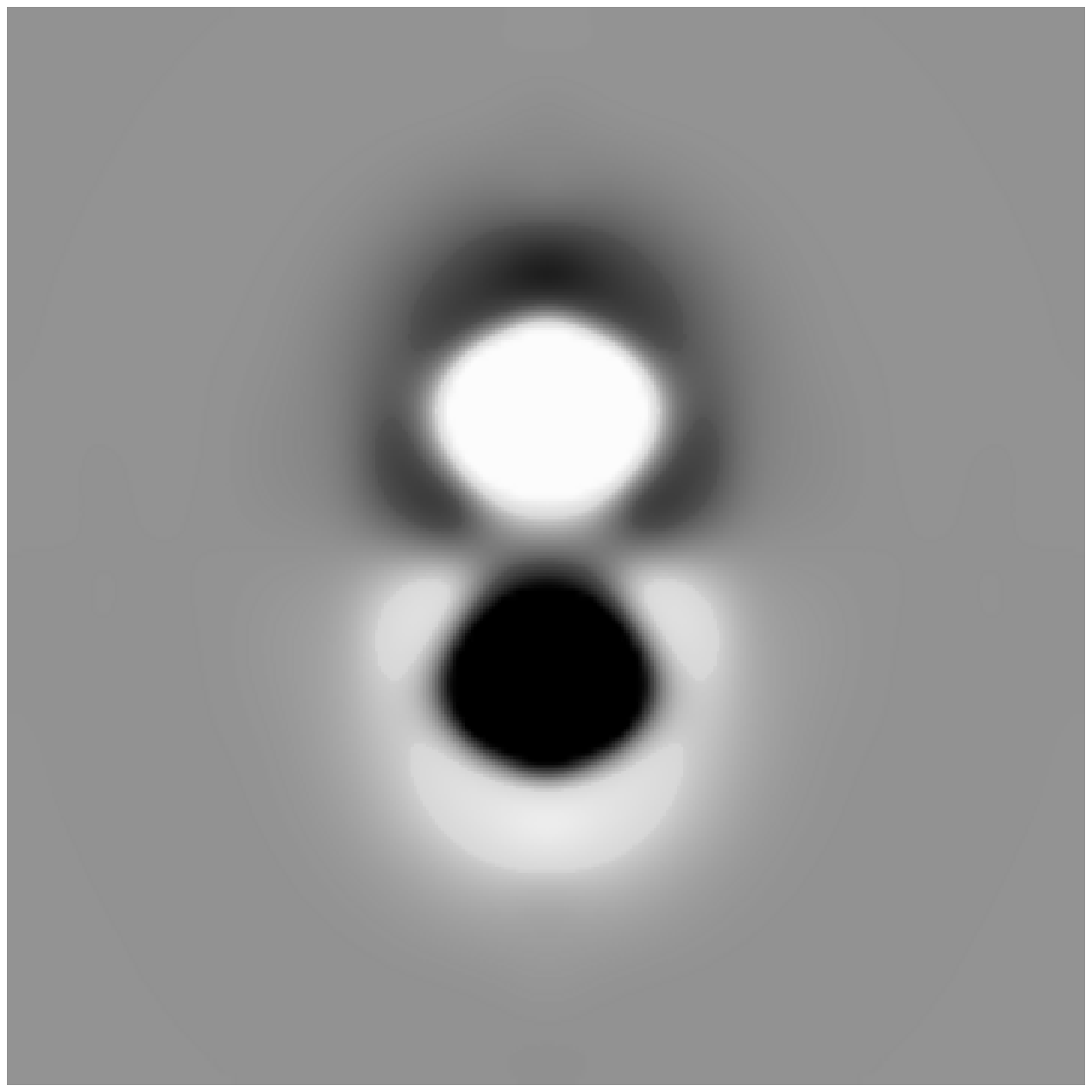}
\includegraphics[height=0.23\textwidth]{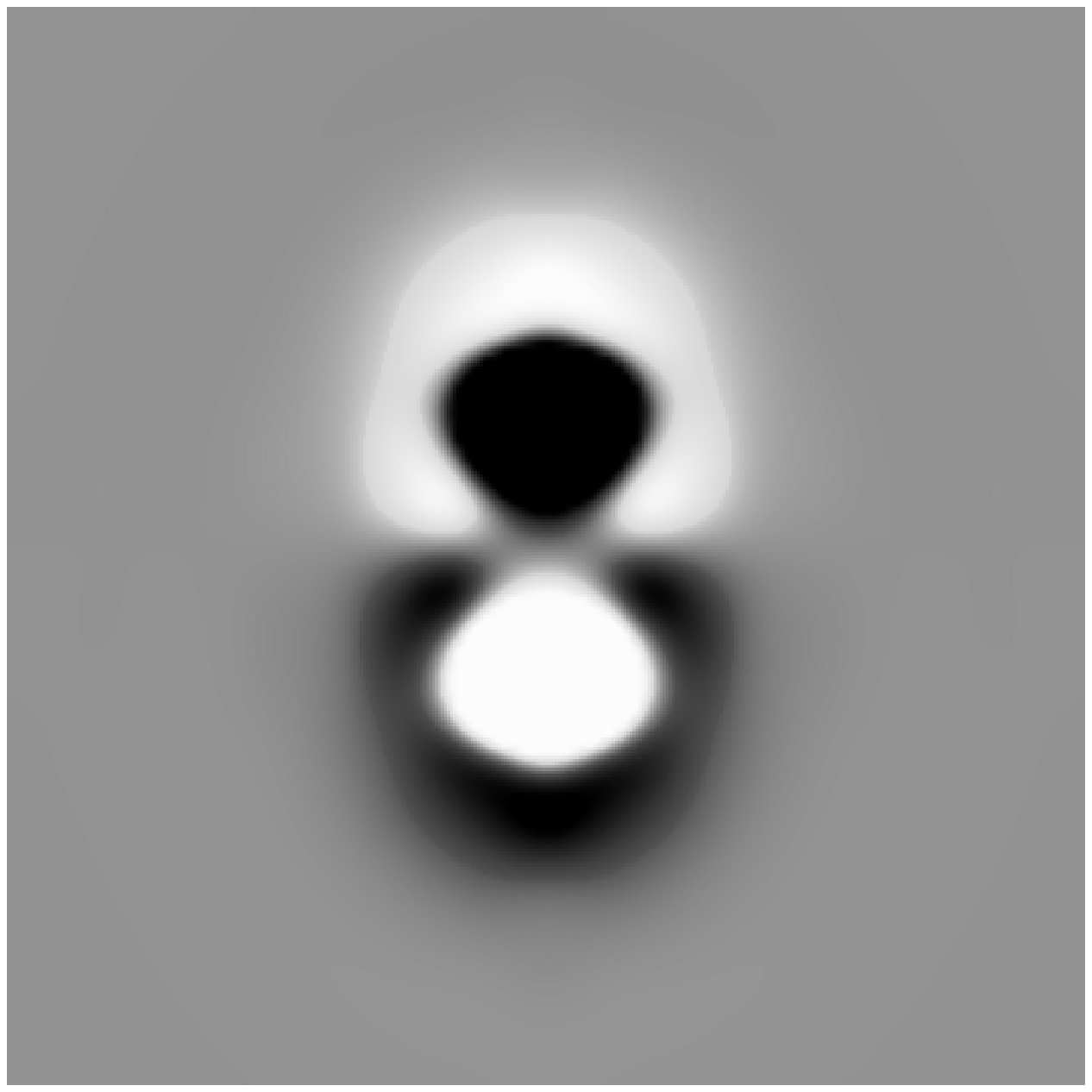}
\includegraphics[height=0.23\textwidth]{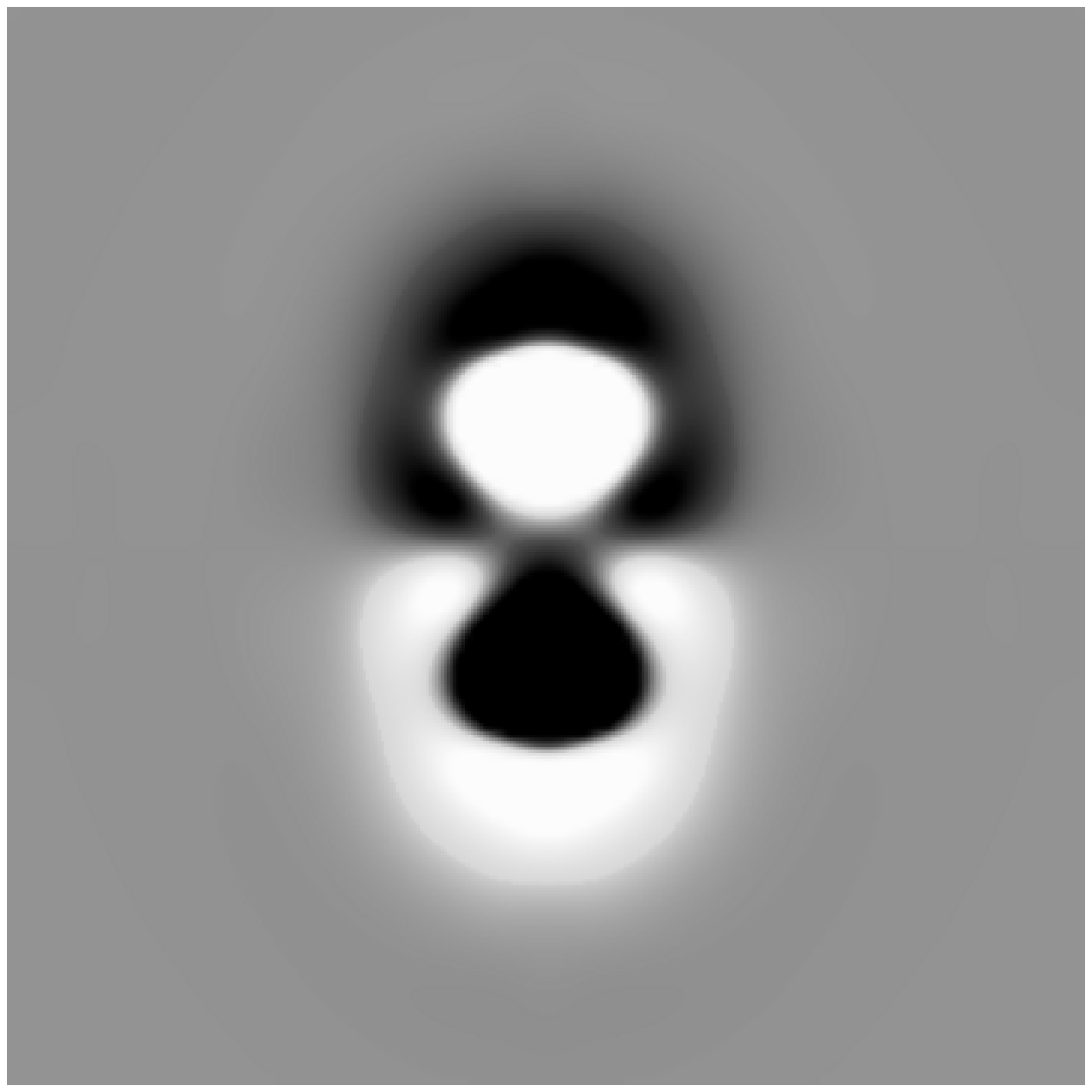}
\includegraphics[height=0.23\textwidth]{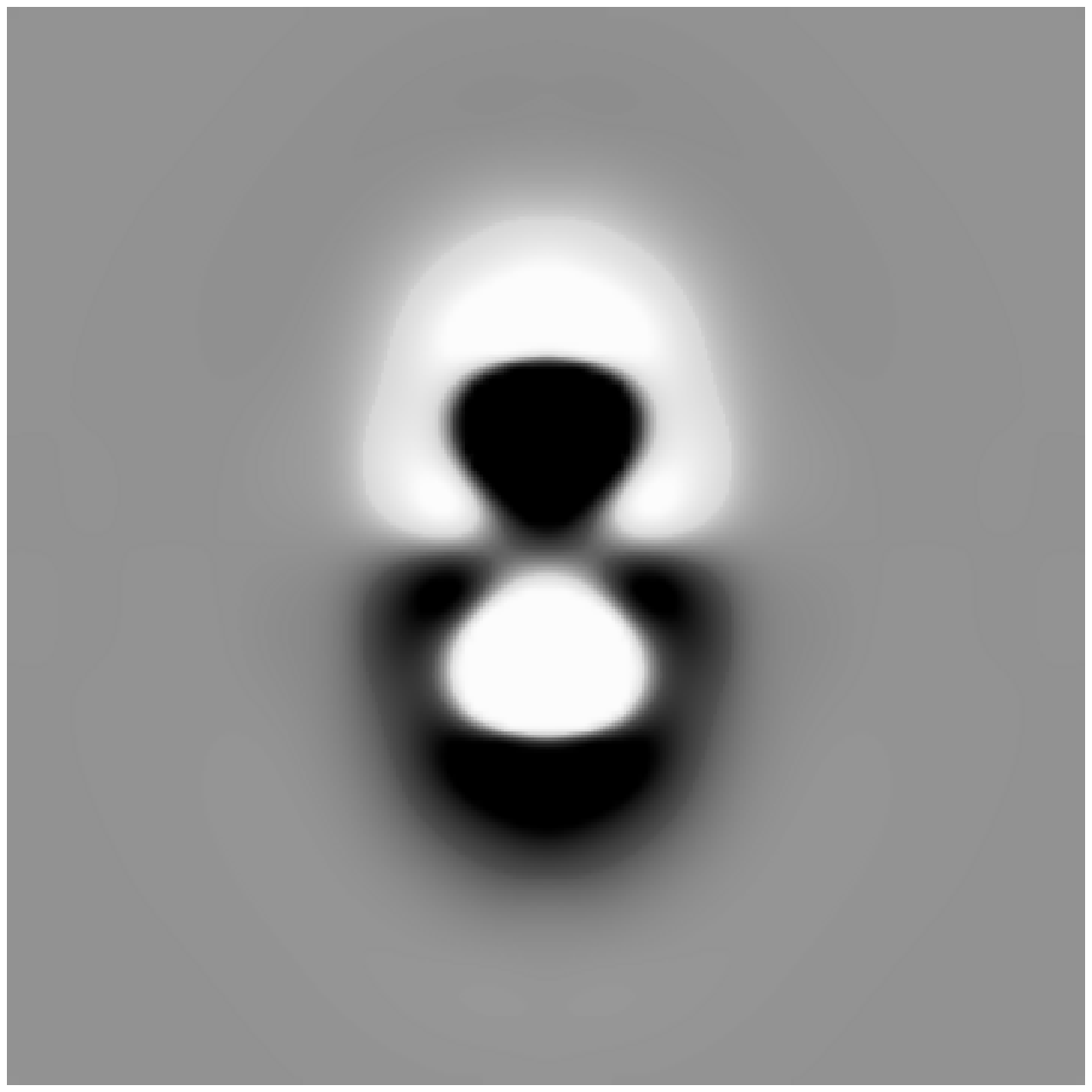}
\begin{minipage}[b]{0.75\textwidth}
\caption{\label{Nakatsukasa}Nuclear TDHF transition densities in $^{14}$Be, see the text.
Reprinted from Ref.~\cite{[Nak07]}, Copyright 2007,
with permission from Elsevier.}
\end{minipage}
\end{center}
\end{minipage}\hspace{0.03\textwidth}%
\begin{minipage}[b]{0.43\textwidth}
\begin{center}
\includegraphics*[width=0.55\textwidth]{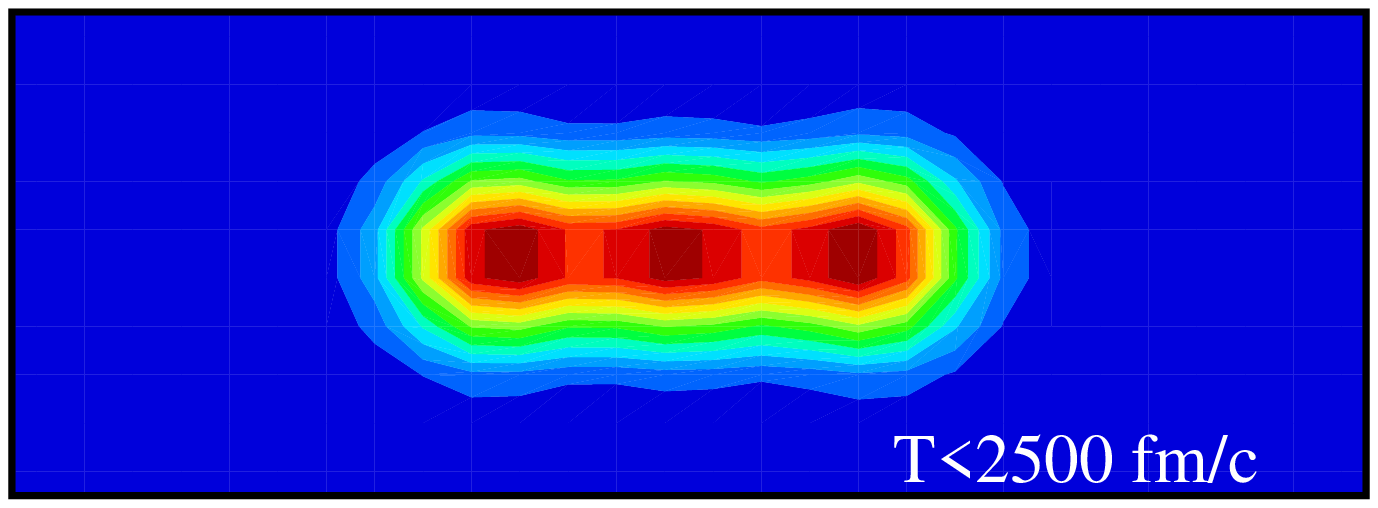}\\[-0.022in]
\includegraphics*[width=0.55\textwidth]{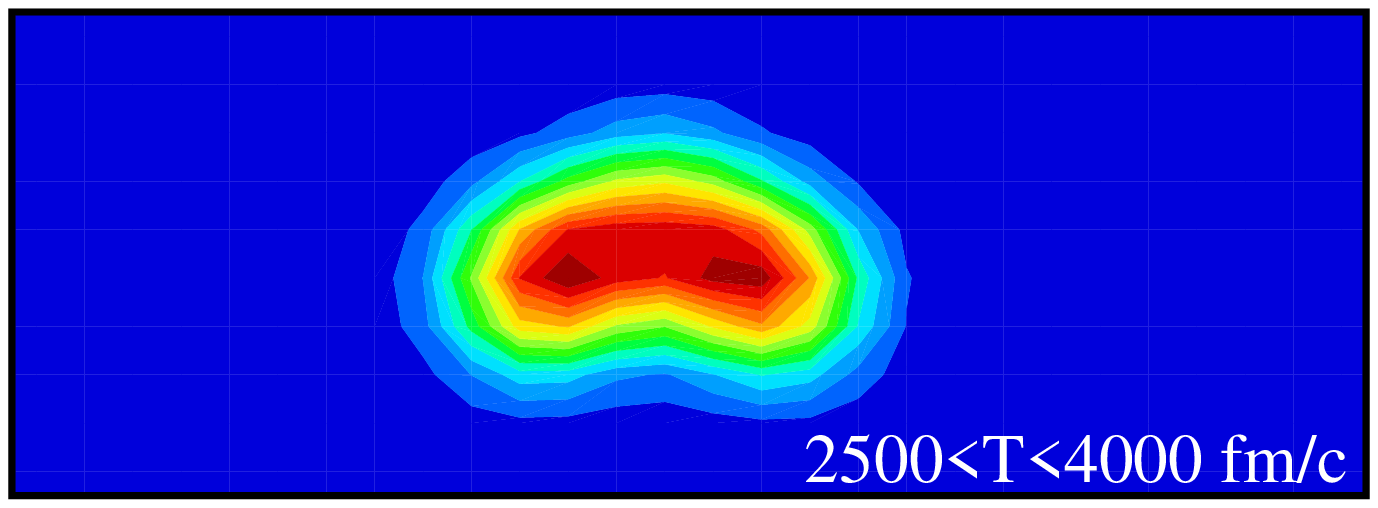}\\[-0.022in]
\includegraphics*[width=0.55\textwidth]{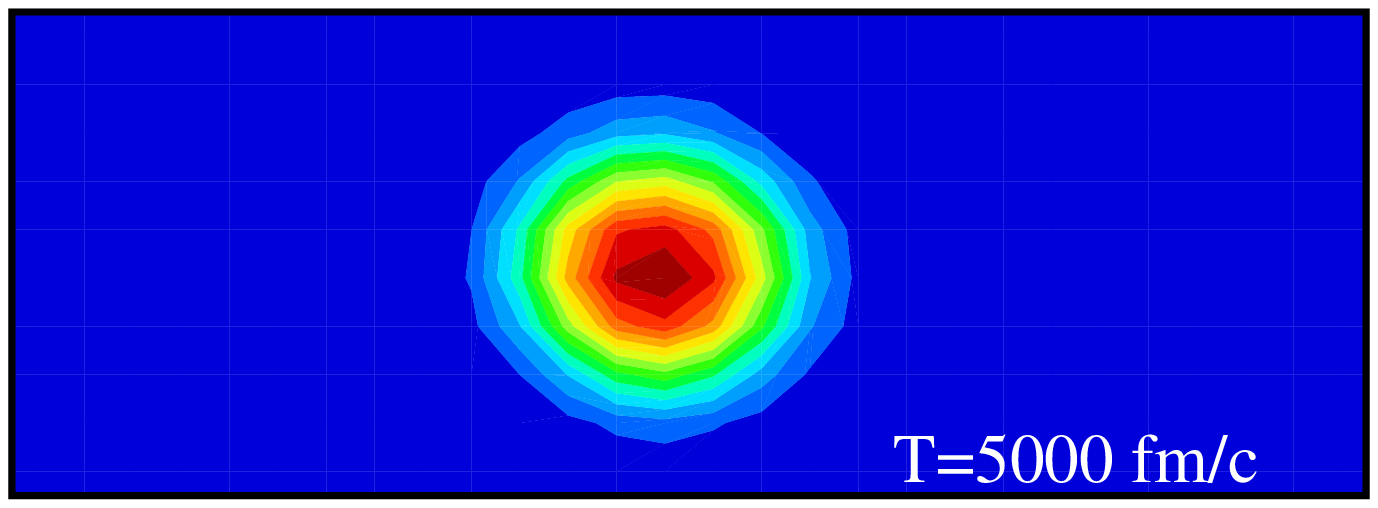}
\caption{\label{Umar}Nuclear TDHF densities in $^{14}$Be, see the text.
Reprinted figure with permission from Ref.~\cite{[Uma10]}.
Copyright 2010 by the American Physical Society.}
\end{center}
\end{minipage}
\end{center}
\end{figure}

The EDF methods were also recently applied within the full 3D
dynamics based on the time-dependent mean-field approach. In
Ref.~\cite{[Nak07]}, the spin-independent transition density was
calculated in the 3D coordinate space for the time-dependent dipole
oscillations. It turned out that one of the Steinwedel-Jensen's
assumptions \cite{[Ste50]},
$\delta\rho_n(\vec{r};t)=-\delta\rho_p(\vec{r};t)$, was approximately
satisfied for $^8$Be. In contrast, in $^{14}$Be, large deviation from
this property was noticed. Figure~\ref{Nakatsukasa} shows how
transition densities $\delta\rho_n(\vec{r};t)$ (lower panels) and
$\delta\rho_p(\vec{r};t)$ (upper panels) evolve in time in the $x$--$z$
plane. The time difference from one panel to the next (from left to
right) roughly corresponds to the half oscillation period. White
(black) regions indicate those of positive (negative) transition
densities. One sees that significant portions of neutrons actually
move in phase with protons.

An interesting 3D EDF time-dependent calculation was recently
performed for the $\alpha$--$^8$Be fusion reaction \cite{[Uma10]}.
Although this calculation aimed at elucidating properties of the
triple-$\alpha$ reaction, it was performed at the energy above the
barrier, where the time-dependent mean-field approach can lead to
fusion, whereas the real triple-$\alpha$ reaction involves tunneling
through the Coulomb barrier. Nevertheless, the studied tip-on initial
configuration in the entrance channel, shown in the upper panel of
Fig.~\ref{Umar}, is probably the preferred one as it must correspond
to the lowest barrier. The calculations lead to the formation of a
metastable linear chain state of three $\alpha$-like clusters which
subsequently made a transition to a lower-energy triangular
$\alpha$-like configuration before acquiring a more compact final
shape, as shown in the lower panels of Fig.~\ref{Umar}.

\section{Conclusions}

The EDF methods are presently very intensely studied and developed in
nuclear physics. Apart from the subjects covered in the present short
review, there is a number of topics that could not be discussed here,
such as: formal aspects of the DFT for self-bound systems; symmetry
breaking effects; {\it ab initio} derivation of the EDF from the
chiral perturbation and Brueckner-Hartree-Fock theories; studies of
weakly-bound systems and the continuum phase-space effects; functionals
non-local in time and the adiabatic connection; new-generation
functionals with higher-order derivatives and/or richer density
dependencies; self-interactions and self-pairing in the EDF; and
ambiguities and inconsistencies when extending the EDF methods to
multi-reference applications.

Moreover, there are several aspects related to the methodology that
were not covered here, such as: the EDF methods based on natural
occupation numbers and/or orbitals; second, extended, and/or
self-consistent (Q)RPA methods; proton-neutron interactions and
isovector terms in the EDF; extrapolations to exotic nuclei and
astrophysical applications; equation of state, symmetry energy, and
neutron stars; relations between functionals describing infinite
systems and finite nuclei; and adjustments of parameters, confidence
intervals, and correlations.

There is also a number of other interesting applications of the
nuclear EDF methods, such as: description of tensor effects and the
spin-orbit splitting; functionals describing pairing correlations;
particle-vibration coupling, single-particle spectra, and widths of
giant resonances; fusion barriers and cross sections within static
and time-dependent calculations; fermion systems in the unitary
regime; neutron skins and pygmy resonances; time-odd terms {\it versus}
spin and orbital M1 resonances, spin-isospin resonances, and
particle-vibration coupling and polarization; incompressibility,
effective mass, and monopole resonances; cluster structures and
models; chirality in rotational bands; di-neutron correlations and
deformation in nuclear halos; and Coulomb frustration effects in the
superheavy nuclei and crust of neutron stars.

In general, the EDF methods provide us with
universal understanding of global low-energy nuclear properties and
feature an impressive array of applications. These methods can be
rooted in the effective-theory approach whereupon the low-energy
phenomena can be successfully modeled without resolving high-energy
properties. Further progress strongly relies upon the use of
high-power computing and faces the challenge of working out a
consistent scheme of consecutive corrections that would allow for the
increased precision and predictive power.

\bigskip

During preparation of this talk, I have received suggestions and
comments from very many of my colleagues; I would
like to thank them very much for their help. In particular, I would
like to thank:
Michael                  Bender,
Karim                    Bennaceur,
George                   Bertsch,
Aurel                    Bulgac,
Rick                     Casten,
Willem                   Dickhoff,
Jerzy                    Dudek,
Nguyen Van               Giai,
Bertrand                 Giraud,
Elvira Moya de           Guerra,
Paul-Henri               Heenen,
Morten                   Hjorth-Jensen,
Pieter Van               Isacker,
Jan                      Kvasil,
Denis                    Lacroix,
Elena                    Litvinova,
J\'er\^ome               Margueron,
Joachim                  Maruhn,
Jie                      Meng,
Witek                    Nazarewicz,
Thomas                   Papenbrock,
Michael                  Pearson,
Jorge                    Piekarewicz,
Nathalie                 Pillet,
Marek                    P{\l}oszajczak,
Paul-Gerhard             Reinhard,
Peter                    Ring,
Wojciech                 Satu{\l}a,
Paul                     Stevenson,
Sait                     Umar,
James                    Vary, and
Dario                    Vretenar.

This work was supported by the Academy of Finland and
University of Jyv\"{a}skyl\"{a} within the FIDIPRO program, by the
Polish Ministry of Science and Higher Education under Contract No.\
N~N~202~328234, and by the U.S.\ Department of Energy under Contract
No.\ DE-FC02-09ER41583 (UNEDF SciDAC Collaboration).

\section*{References}
\bibliographystyle{iopart-num}
\providecommand{\newblock}{}

\end{document}